\documentclass[prl,onecolumn,superscriptaddress,showpacs,preprintnumbers,amsmath,amssymb,floatfix,preprint]{revtex4-1}
\usepackage{graphicx}
\usepackage{hyperref}
\usepackage{epsfig}
\usepackage{color}
\usepackage{soul}

\newcommand{\BT}{Bi$_2$Te$_3$ }

\begin{document}

\title{Joule meets van der Waals: Mechanical dissipation via image potential states on a topological insulator surface}
\author{D. Yildiz}
\email[Correspondance and requests for materials should be addressed to D.Y. (email:
dilek.yildiz@unibas.ch) or M.K. (email: marcin.kisiel@unibas.ch)]{}
\affiliation{Department of Physics, University of Basel, Klingelbergstrasse 82, 4056 Basel, Switzerland}
\author{M. Kisiel}
\email[ Correspondance and requests for materials should be addressed to D.Y. (email:
dilek.yildiz@unibas.ch) or M.K. (email: marcin.kisiel@unibas.ch)]{}
\affiliation{Department of Physics, University of Basel, Klingelbergstrasse 82, 4056 Basel, Switzerland}
\author{U. Gysin}
\affiliation{Department of Physics, University of Basel, Klingelbergstrasse 82, 4056 Basel, Switzerland}
\author{O. G\"{u}rl\"{u}}
\affiliation{Department of Physics, Istanbul Technical University, Maslak, 34469, Istanbul, Turkey}
\author{E. Meyer}
\affiliation{Department of Physics, University of Basel, Klingelbergstrasse 82, 4056 Basel, Switzerland}


\begin{abstract}

Dissipation mechanisms are experimentally studied on topological insulator surfaces of Bi$_2$Te$_3$, where common Joule dissipation was observed to be suppressed due to topologically protected surface states. Thus, a novel type of dissipation mechanism is observed by pendulum AFM, which is related to single electron tunneling resonances into image potential states that are slightly above the \BT surface. The application of a magnetic field leads to the break down of the topological protection of the surface states and restores the expected Joule dissipation process. Nanomechanical energy dissipation experienced by the cantilever of pendulum AFM provides a novel source of information on the dissipative nature of the quantum-tunneling phenomena on the topological insulator surface. 
 
\end{abstract}

\maketitle

\section{Introduction}

Topological insulators (TI) attract great attention due to the potential use of the topologically protected surface electronic states in advanced communications and information processing systems, as well as in quantum computing \cite{Chen2009}. The layered compound \BT is a model TI with prevented electron back-scattering, long electron lifetimes \cite{Hasan2010, Seo2010, Zhang2009B} and reduced electrical resistance at low temperatures due to the effect of weak anti-localization \cite{Hong2011}. Although electronic properties of topological insulators have been studied extensively, the frictional response of their surfaces is yet to be reported. The impact of electronic structure and topologically protected surface states on the dissipative interaction between an oscillating tip and the sample is the scope of the present study.

Image potential states (IPS) on metallic surfaces \cite{Himpsel1986,Dose1987,Echenique1991,Berthold2002,Wahl2003,Schouteden2009,Niesner2014C} resembling Rydberg series were observed on several topological insulators \cite{Sobota2012,Sobota2013,Niesner2012,Niesner2014,Niesner2014B}, with the energy states lying slightly below the vacuum level. Angle-resolved two-photon photoemission (2PPE) studies of Bi$_2$Te$_2$Se surfaces reported on the first IPS to be  at $E=4.5 \mathrm{eV}$ above Fermi level \cite{Niesner2012}. IPS are weakly coupled to the bulk in comparison to the other surface states. The image potential states of TIs have relatively long lifetimes in the order of $\mathrm{fs}$, comparable to metallic surfaces \cite{Niesner2014C}. In Scanning Tunneling Spectroscopy (STS), IPS are detected as Gundlach oscillations, which is a phenomenon of field emission resonance through IPS in the tip-sample gap \cite{Gundlach1966}. The IPS are located a few nm away from the surface with an increasing tendency for higher quantum numbers n. The wave functions of IPS were reported to be extended up to 20nm away from the surface in two photon photoemission experiments \cite{Hofer1997}. Although the presence of such IPS is well known, their impact on non-contact energy dissipation is not explored, so far. 

Atomic force microscope (AFM) utilising a cantilever oscillating like a tiny pendulum over a surface is designed to measure extremely small non-contact energy dissipation and serve as an ultra-sensitive, non-invasive spectroscopy method \cite{Kisiel2011, Langer2013, Kisiel2015} (see Supplementary Information section S1 and Supplementary Figure 1). It has been shown that non-contact pendulum geometry AFM (pAFM) is sensitive to different types of energy loss mechanism in non-contact regime, where the oscillating probe is separated from the sample by a vacuum gap. In particular, phonon excitation \cite{Kisiel2011}, Joule ohmic dissipation \cite{Kisiel2011} or van der Waals dissipation \cite{Stipe2001,volokitin2007} were reported. 

Here we combine Scanning Tunneling Spectroscopy (STS) with pAFM on a clean \BT surface (see Methods). The measurement setup is described in Figure 1(a). Rydberg-like series of conductance maxima are observed by z-V spectroscopy, where the bias is swept with an active feedback in constant current mode. Thus field emission resonances are very well resolved up to the fifth order.


Mechanical dissipation measurements by pAFM show increased energy losses at discrete separations and voltages up to distances of $14\mathrm{nm}$. Combined STM/pAFM measurements reveal that the Gundlach oscillations are accompanied by increased mechanical dissipation.   Therefore, we interpret the enhanced dissipation losses at discrete separations and voltages to charge fluctuations of the IPS. Tunneling processes lead to occupancy and de-occupancy of the IPS, which is detected by pAFM.  If magnetic fields are applied, we do observe that Joule-type dissipation rises, which is presumably related to the destruction of the topological protection, which opens the channel for scattering to bulk states giving rise to increased Joule dissipation as it is more common on ordinary metallic surfaces \cite{Kisiel2011}.

\section*{Results}

\subsection{Scanning Tunneling Microscopy imaging and bias dependent tunneling spectra}

A typical Scanning Tunneling Spectroscopy (STS) spectrum taken at close tip-sample distances at 5K is presented in Figure 1(b). The inset shows an atomically resolved topography image, acquired in constant current mode STM performed with a gold-coated cantilever tip (see Methods for details of STM and STS measurements).  Close to Fermi energy, the $dI/dV$ spectrum reveals a linear dependence on bias voltage and the linear part of the curve crosses the voltage axis at about $V_s=-0.3V$ bias voltage. Depending on the crystal growth conditions and doping, values between -0.1V to -0.4V have been reported \cite{Neupane2012, Miyamoto2012, Niesner2012}. The similar linear density of states, resembling a Dirac cone, is a signature of the topologically protected surface state of pristine \BT \cite{Schouteden2016}. It is, therefore, reasonable to assume that the topologically protected electronic structure of the \BT surface is preserved \cite{Schouteden2016}. 

\begin{figure}[!ht]\centering
\includegraphics[width=0.65\textwidth]{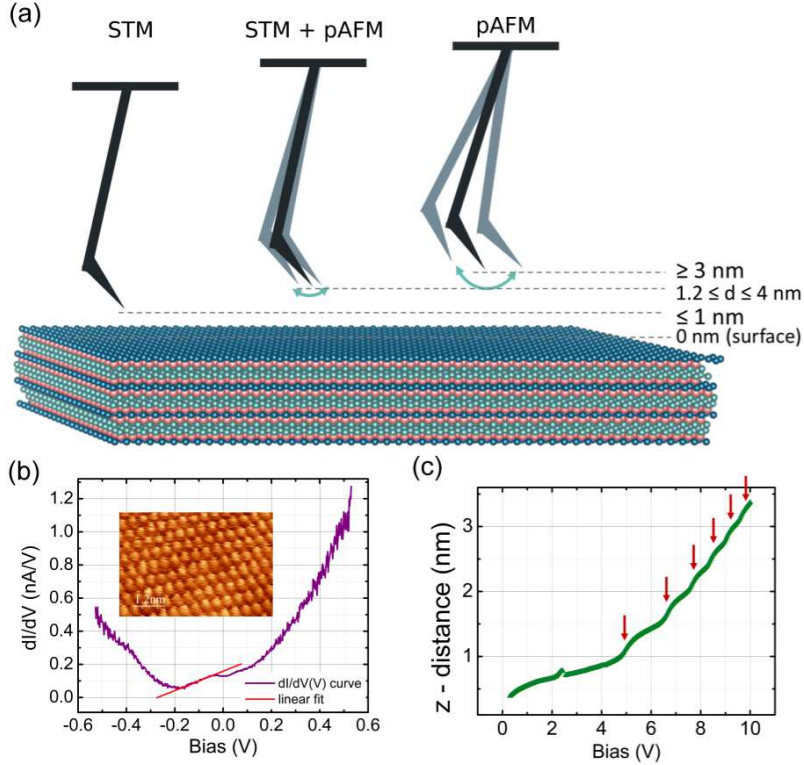}
\caption{\label{fig_stm} 
\textbf{STM and STS measurements on the \BT surface performed at T=5K.}
a) Measurement protocol showing STM and AFM operating in pendulum geometry. The STM operates at the closest tip-sample distance, while the pAFM nominal working distance is typically larger than 3nm, b) STS showing the Dirac-like cone together with STM topography (inset). $dI/dV$ was numerically calculated from an I(V) curve and shows the Dirac-like cone. Tunneling parameters: $\mathrm{I_t}= 80\mathrm{pA}, \mathrm{V_s}= 300\mathrm{mV}$.  c) z(V) curve showing series of image potential states confined between the STM tip and \BT surface. The feature appearing close to 2V may originate from surface/subsurface or interlayer defect states \cite{Pivetta2005}. Steps present above 5V are the image potential states of the crystal that are measured above the surface.}
\end{figure}

$z(V)$ spectroscopy measurements were performed by a continuous sweep of the tip-sample voltage while keeping the current constant by the STM feedback. Thus, the STM tip retracts if there is an increase in the tunneling current, thereby revealing the Rydberg-like series of electronic states as shown in Figure 1(c). The total change of tip-sample distance z observed between the $V_s=1V-10V$ voltage sweeps is about 2.5nm, showing 6 step-like increments. The change of each $\Delta$z step is about 300pm. The first peak (n=0) located close to $V_s=5V$ is related to the local work function \cite{Wahl2003} of the surface. $z(V)$ spectra show a sequence of field emission resonances numbered by the quantum numbers, n=0,1,2,3,4,5, which are visible in the differentiated $z(V)$ curves as shown in Figure 2(a). Recent 2PPE experiments on bismuth rich surfaces reported IPS \cite{Sobota2012, Sobota2013, Niesner2012, Niesner2014, Niesner2014B}, and apart from Rydberg-like series, a peak localized at 2.5eV energy, which is present in our STS data as well. In local probe measurements its presence is location dependent and thus it might be related to the subsurface defect or interlayer/interface states (see Supplementary Figure 2 for location dependent experiments and local $z(V)$ spectra) \cite{Pivetta2005, Bose2010}. It needs to be mentioned that, since the STM measurements are performed with a tip mounted on a cantilever, the static deflection of the sensor was monitored. Forces in the range of $pN$ (see Supplementary Figure 3) were detected.

\subsection{Field emission resonances probed by oscillating STM tip}

We performed combined STM and AFM based spectroscopy measurements, by means of $z(V)$ spectra obtained from pAFM running in STM mode, while the tip was oscillating. Energy dissipation of the cantilever was monitored, while simultaneously measuring the tunneling current between the tip and the \BT surface. The amplitude of lateral oscillations during combined STM/AFM spectroscopy measurements by pAFM was set to $  \sim30 \mathrm{pm}$. Due to the geometry of the tip, this results in amplitude normal to the surface of $2-3\mathrm{pm}$ (see Supplementary Figure 4). This is two orders of magnitude smaller than the change of tip-sample distance $\Delta$z observed in $z(V)$ curves. In Figure 2(a) $dz/dV$ data show the IPS related resonances for static (black) and oscillating (red) STM tip.  

\begin{figure}[!ht]\centering
\includegraphics[width=0.9\textwidth]{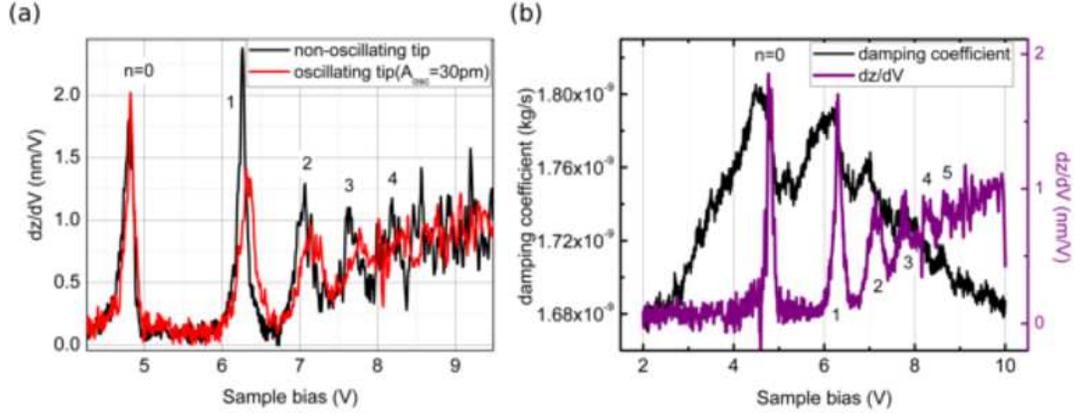}
\caption{\label{fig_oscstm} 
\textbf{Series of field emission resonances measured with a static and oscillating STM tip, and simultaneous energy dissipation measurement at T=5K.}
a) $dz/dV$ measurements performed at the same surface spot with static (black) and oscillating (red) STM tip. The amplitude of lateral oscillation was 30 pm, and n enumerates the image potential states. The oscillating tip causes a broadening of the IPS peaks that are shown in dz/dV(V) curve at $V_s>5\mathrm{V}$, while n=0 IPS at $V_{s}=4.8\mathrm{V}$ is unchanged in dynamic measurements as compared to static situation b) simultaneous  STS-$dz/dV$ (violet) and AFM-dissipation (black) measurements performed by oscillating tip. The field emission resonances observed in $dz/dV$ spectra are accompanied by changes in the dissipation spectrum.}
\end{figure}

IPS at bias voltages $V_s>5\mathrm{V}$ are broadened by a factor of 2, while n=0 IPS at $V_{s}=4.8\mathrm{V}$ is almost unchanged in dynamic measurements compared to the static case (see Supplementary Figure 5 for further discussion on the Full With and Half Maxima (FWHM) of the IPS). In both cases, the FWHM indicates that IPS on \BT surface are relatively long lived, with the lifetime $\tau \propto \mathrm{fs}$, in agreement with reported literature values \cite{Niesner2014}. The tip oscillation smears out the IPS for $\mathrm{n>4}$, presumably due to reduced sensitivity at far distances and intermixing of the states with high n by the tip induced oscillating tunneling barrier. Oscillating tip STS measurements are used to perform simultaneous STM/STS and pAFM dissipation spectroscopy measurements, as shown in Figure 2(b), where the series of IPS is accompanied by changes in dissipation signal. Here, four IPS are visible. The dissipation signal rises for each quantum number. The drop of the dissipation towards the maximum of the $dz/dV$ curves is related to the z retraction of the z(V) spectroscopy. Although the frictional response of the AFM is known to depend on tip-sample distance and bias voltage that is applied between the tip and the sample \cite{Kisiel2011, Stipe2001}, the simultaneous increase of the dissipation signal and the correspondence to the series of IPS provides strong evidence that both phenomena are linked together, and the field emission resonances affect the mechanical nano-dissipation on \BT surface. 

\subsection{Dissipation spectroscopy measurements by pAFM}

Apart from provoking conventional forms of energy dissipation mechanisms, such as phonon and Joule losses \cite{Volokitin2006, Kisiel2011}, the external perturbation caused by an oscillating tip might push a finite quantum system towards a transition or a level crossing with subsequent fluctuation and relaxation of the system, eventually resulting in the enhancement of energy loss \cite{Cockins2010, Kisiel2018, Langer2013}. On \BT surface, we claim that the energy losses occur when the oscillating tip couples to the charge fluctuations of IPS due to electron tunneling. In the AFM mode, the tip is retracted away from the STM operation distance and the feedback is switched from STM to AFM operation. The tip is oscillated with $300-400\mathrm{pm}$ lateral oscillation amplitude and the oscillations perpendicular to the sample are in the order of $30 \mathrm{pm}$. Before measuring dissipation, the tip-sample distance and oscillation amplitude are controlled in order to exclude modulation currents due to the cantilever oscillation. After retraction, the sample bias was swept between 10V to -10V while the tip is grounded and dissipation and frequency shift spectra are recorded. The pendulum AFM voltage dependent measurements show parabolic dependence of the frequency shift ($\Delta f$) and non-monotonic dissipation, as shown in Figure 3(a) (see Methods section for details about $\Delta f$ and dissipation signals). At 5nm tip-sample separation, we observe the first peak in dissipation data located at $V=\pm7.4\mathrm{V}$ and a second peak at $V=\pm 8.5\mathrm{V}$. Both are symmetric with respect to contact potential difference (CPD) (see Supplementary Information section S3). This is in analogy to AFM measurements of weakly coupled quantum dots \cite{Stomp2005} or molecules in break junctions \cite{Paulsson2002}, where the voltage drop is divided across two capacitances (tip-molecule capacitance and molecule-sample capacitance). If the capacitances are comparable, symmetric case is observed. The $\Delta f$ signal acquired simultaneously with dissipation signal shows deviations from the simple parabolic dependence (see Supplementary Figure 6) which coincides with the position of the enhanced dissipation, meaning after each dissipation peak, cantilever tip-sample capacitive coupling changes and the tip is subject to slightly different force fields.

\begin{figure}[!ht]\centering
\includegraphics[width=0.9\textwidth]{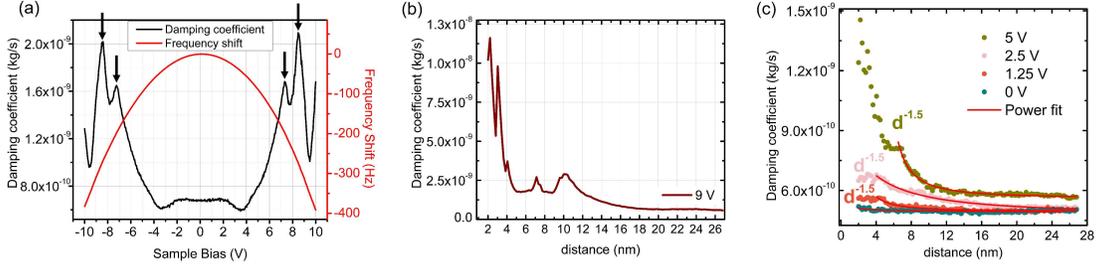}
\caption{\label{fig_oscstm} 
\textbf{Voltage and distance dependent AFM  dissipation measured at T=5K on \BT surface.}
a) Tip-sample voltage dependent dissipation (black) taken at constant tip-sample distance, d=5nm, shows series of dissipation enhanced features marked by arrows positioned symmetrically with respect to bias voltage $V_{s}=0\mathrm{V}$ and located around $V_{s}=\pm 7.4\mathrm{V}$ and $V_s=\pm 8.5\mathrm{V}.$ Simultaneously with dissipation the frequency shift signal (red) was acquired. The lateral oscillation amplitude was equal to 300pm ($f_0=269\mathrm{kHz}, k=58 \mathrm{N/m}$.) b) Dissipation versus distance spectrum taken at sample bias $V_s=9\mathrm{V}$. The dissipation peaks after subsequent opening of each dissipation channel are observed as a rise of dissipation plateau for distances smaller than 10nm, c) Dissipation spectra measured at sample bias $V_s \leqslant 5\mathrm{V} $ follow $d^{-3/2}$ power law in agreement with theory of non-contact dissipation on thin metallic film on an insulator \cite{gnecco_meyer_2007}. On b and c the lateral amplitude of cantilever oscillation was  $400\mathrm{pm}$.}
\end{figure}

The distance dependence of energy dissipation is shown in Figure 3(b). The data were obtained by approaching the tip towards the \BT sample with a constant voltage of $V_s=9\mathrm{V}$. Two main features are present in the dissipation versus distance spectra: Firstly, a series of dissipation peaks at $z=2,6,10\mathrm{nm}$ distances are observed. Secondly, we notice an overall rise of dissipation plateau after tip approaches to the first dissipation peak. At distances larger than $z>10\mathrm{nm}$ the minimum value of damping coefficient is equal to $\Gamma=7.0 \cdot 10^{-10} \mathrm{kg/s}$ and then levels off to be about $\Gamma=2 \cdot 10^{-9} \mathrm{kg/s}$ at closer tip-sample distance. This rise of dissipation plateau after the first dissipation peak suggests the opening of a specific dissipation channel at distances closer than $z=9\mathrm{nm}$ and tip-sample voltage $V_s=9\mathrm{V}$. Distance - dependent dissipation spectra measured at sample bias $V_s \leqslant 5\mathrm{V} $ are shown in Figure 3(c). The spectra show $d^{-3/2}$ power law in agreement with the theory of non-contact dissipation on thin metallic film on an insulator \cite{volokitin2007, gnecco_meyer_2007}.

\begin{figure}[!ht]\centering
\includegraphics[width=0.6\textwidth]{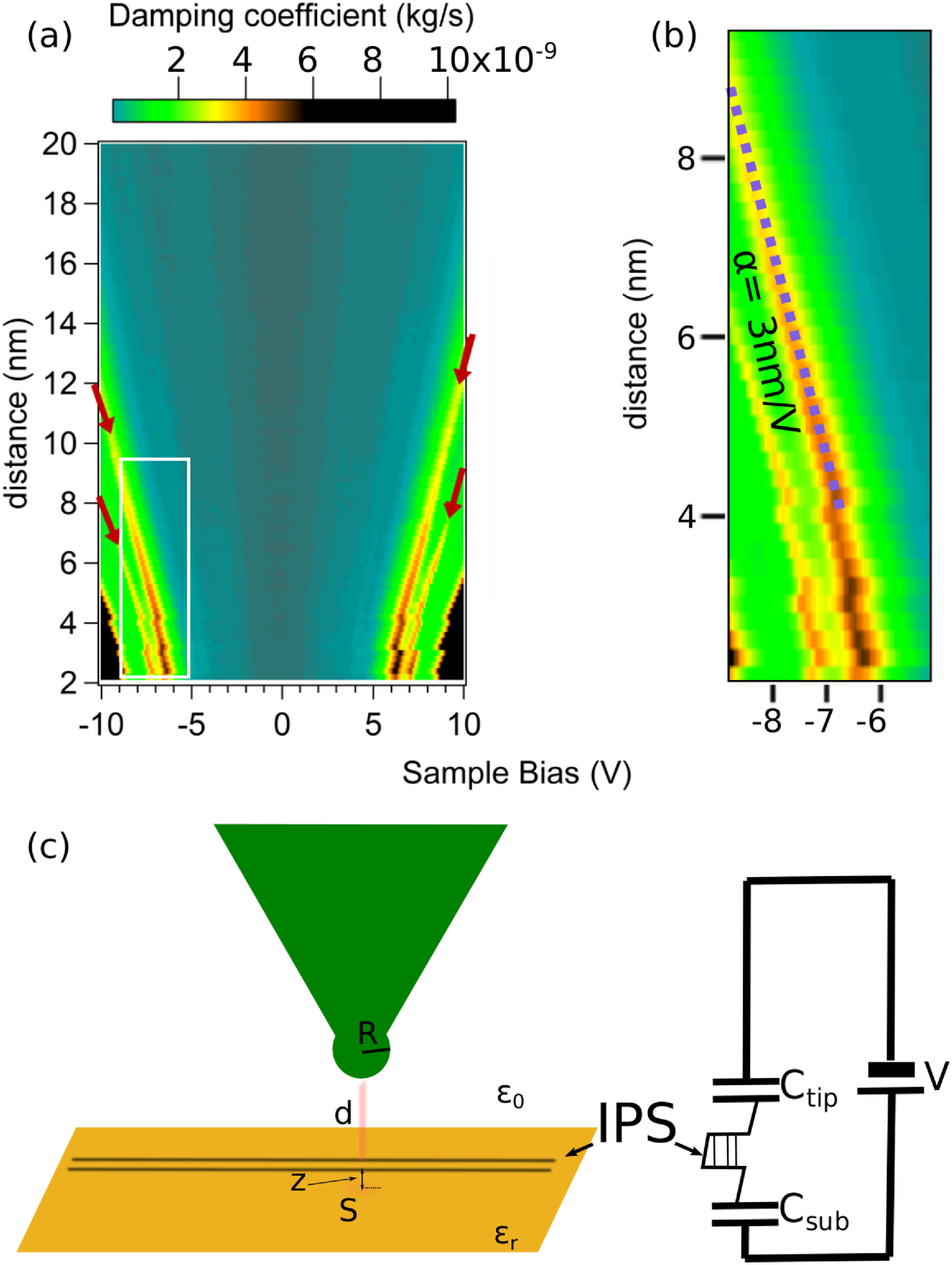}
\caption{\label{fig_oscstm} 
\textbf{Energy dissipation map on \BT plotted versus distance and tip-sample voltage.}
Darker contrast represents four large dissipation features marked by red arrows. The lateral amplitude of the cantilever oscillation was  $400\mathrm{pm}$. The white rectangle marks the region shown in b, b) shows the detailed zoom of the dissipation map for bias voltage between $-9V<V_s<-5V$. Two dissipation features are visible, and for distances larger than 4nm they move to larger voltages due to a decrease of capacitive coupling. The nonlinearity seen at distances $d<4\mathrm{nm}$  suggests that the tip radius is equal to about 4nm. The schematics of the tip-sample geometry and the equivalent electrical circuit is shown in c, where $R$, $d$, $z$ and $S$ stands for tip radius, tip distance to the IPS, the distance of IPS to the surface and tip projected "active" surface area, respectively. In analogy to break junction geometry the voltage drop is divided across two capacitances: tip-IPS $(C_{tip})$ and IPS-\BT surface $(C_{sub})$.}
\end{figure}

The dissipation map in  Figure 4(a) shows the distance and voltage dependence of the damping coefficient $\Gamma(V,z)$ of the cantilever. Red arrows mark the positions of the dissipation peaks on the map. The maxima are observed at non-zero biases even at close distances, which indicates that dissipation is not force but voltage controlled. It has to be noted that the van der Waals force present at lower biases cannot cause the discussed dissipation features. Similar to the case of charging of quantum dots \cite{Cockins2010, Kisiel2018}, the amount of dissipated energy is also in the order of tens of meV per cycle indicating a single electron tunneling process. The position of dissipation peaks shifts linearly towards higher bias voltages with increasing tip-sample distance due to the decrease of capacitive coupling between tip and sample. This is shown in detail in Figure 4(b), and the measurement reported on lever arm $\alpha = 3nm/V$. Thus, at far distances, the voltages of dissipation features are shifted compared to the voltages observed by STM, as shown in Figure 2. This can be understood by taking into account that AFM data are influenced by the voltage drop across the vacuum gap, which is divided by the two effective capacitances $C_{tip}$ and $C_{sub}$ (see Figure 4(c)). 
At very close distances below 4 nm, we observe a nonlinearity in tip-sample capacitive coupling. This suggests that the tip radius $R$ is approximately equal to 4nm \cite{Sarid1991}. The extrapolation of the first dissipation maximum to the $ d\leq 1\mathrm{nm}$, a working distance of STM, results in comparable energy scale seen by STS.

\subsection{Dissipation measurements under magnetic field}

To further examine the effect of magnetic field on the dissipation and corroborate on the effect of weak anti-localization  \cite{Hong2011, Sessi2014} we performed the dissipation measurements under external magnetic fields ranging from $B=0-0.8 \mathrm{T}$ oriented perpendicularly to the sample surface (see Figure 4(a)). The tip was positioned at a 5 nm distance above the surface.  As the magnetic field rises, dissipation maxima become less pronounced, and the overall dissipation background raises as marked by green arrows in Figure 4(a). The spectrum obtained for $B=0.8\mathrm{T}$ resembles the common Joule dissipation parabolic shape obtained on ordinary metal surfaces \cite{Kisiel2011}. Moreover, we noticed the rise of the overall dissipation background even for compensated CPD voltage $(V_{CPD}=0V)$ as shown in Figure 4(b). Thus, we conclude that Joule dissipation, connected to bulk connectivity, rises for a magnetic field $B>0.2\mathrm{T}$, where the spin-momentum locking appears to be destroyed, and back-scattering becomes prominent.  According to Kohler's rule $R(B)/R(0) \approx 1+(\mu B)^2$  \cite{Hong2011,olsen_1962} the metallic sample resistivity in the weak magnetic field limit exhibits a $B^2$ dependence, where $\mu$ is the mobility of the film.  Since the dissipated power is proportional to the magnetoresistance of the sample (see Supplementary Information section S2 for the relation of measured AFM power dissipation to the sample resistivity), the dissipation curve should show a parabolic dependence on $B$. The parabolic fit of the data for $B>0.2\mathrm{T}$ is shown as a solid red line. Accordingly, we conclude that AFM dissipation is sensitive to the effect of weak anti-localization, the unique property of the topological matter and the suppression of Joule type of dissipation on topologically protected surfaces is crucial for observation of dissipation due to the presence of image potential states.

\begin{figure}[!ht]\centering
\includegraphics[width=0.9\textwidth]{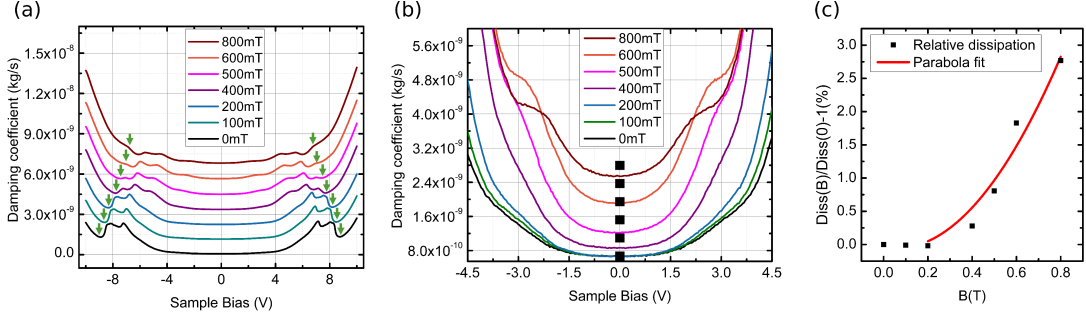}
\caption{\label{fig_versusdf} 
\textbf{Energy  dissipation as a function of the magnetic field measured at T=5K on \BT surface}. (a) shows dissipation versus bias spectra for different external magnetic fields. The tip is at a constant distance (z=5nm from the tunnel distance). The black curve with well-pronounced dissipation peaks is for $B=0 \mathrm{T}$. As the magnetic field rises, overall Joule type of dissipation increases and dissipation features gets suppressed. This effect is best visible in the proximity of the dissipation peaks as marked by arrows. A constant vertical shift is applied to curves for visibility. (b) Dissipation versus voltage for different magnetic fields. No vertical offset was applied to the curves. The black dots indicate the data shown in (c) where the relative dissipation versus magnetic field is shown for $V_s=V_{CPD}$. The strong rise of dissipated power above B=0.2T is related to the breaking of topological protection of the surface state by the $B$ field. The red curve is a parabolic $B^2$ fit to the data. 
}
\end{figure}

\section{Discussion}

To corroborate onto the origin of observed energy dissipation we estimate the damping coefficient following theoretical predictions for Joule dissipation as given by Volokitin et al. \cite{gnecco_meyer_2007} and formula (19.73) therein:

\begin{equation}
    \Gamma = \frac{(4\pi \epsilon_0)^2 w(V^2+V_0^2)R^{1/2}}{2^{9/2} \pi \sigma d_f d^{3/2}}
\end{equation}

The theoretical model considers a metallic film on top of an insulating/semiconducting bulk substrate. Such a model accounts for the topologically protected electronic structure of the sample and the measured dissipation versus distance (see Figure 3c) indeed follows $\Gamma \propto d^{-3/2}$ dependence. A more detailed analysis with this model seems not adequate because of the lack of knowledge of the parameters for the case of TI.  

In Figure 4 (a,b) the dissipation maxima shifts with voltage and distance, due to voltage division between $C_{tip}$ and $C_{sub}$ as shown in Figure 4c. The symmetry of the curves is in analogy to nc-AFM measurements of quantum dots \cite{Stomp2005} and molecules on thick insulators \cite{Fatayer2018}. The symmetric appearance is also common in break junction experiments, where resonant tunneling is observed at both polarities \cite{Paulsson2002}. At the voltages, where dissipation maxima occur, we do observe small irregularities in the $\Delta f$ signal which fit well two capacitor model with different values for $C_{tip}=0.17aF$ and $C_{sub}=0.8aF$ (see Supplementary Figure 6) at $d=5nm$ tip-sample distance. The ratio $C_{tip}/C_{sub}=0.2$ gives the position of IPS above the \BT surface equal to $z=0.4 nm$, which is a realistic estimate. Moreover, experiments with different tip material show that the positions of dissipation peaks and related $\Delta f$ are symmetric with respect to CPD (see Supplementary Fig. 8), as expected for a two-capacitor model.

\begin{figure}[!ht]\centering
\includegraphics[width=0.6\textwidth]{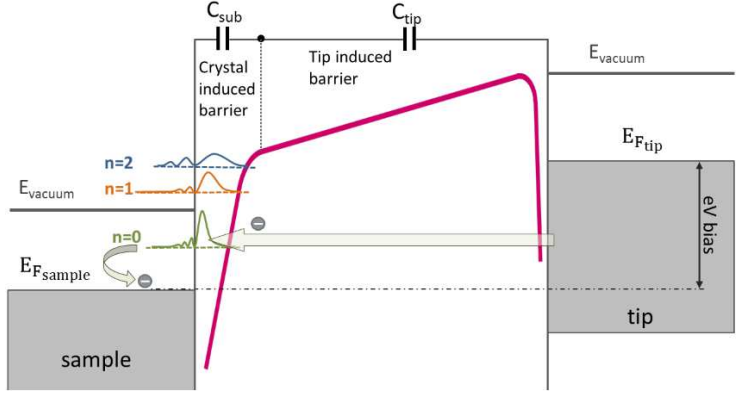}
\caption{\label{fig_imstat_disc} 
\textbf{Single electron tunneling between IPS of \BT sample and an oscillating AFM tip.} With positive bias voltage applied to the sample, the conditions for single/few electron resonant tunneling is created between tip and sample through the IPS at the gap. 
The tunneling process increases the charge fluctuations in the system and thus enhances AFM dissipation. The electric field applied between tip and sample is in the order of $E \backsim 10^9 Vm^{-1}$, and the tip-IPS and IPS-sample capacitances are marked $C_{tip}$ and $C_{sub}$ respecitvely.} 
\end{figure}

The observed dissipation features phenomenology fits the model (Figure 6) of resonance tunneling on \BT surface as follows. For a specific voltage and distance, the tunneling leads to the occupation of the IPS. Electrons tunnel to the IPS continuously and the decay time of occupied IPS is in the femtosecond range. The IPS charging and subsequent charge relaxation lead to substantial charge fluctuations in the system and thus give rise to the increase of mechanical dissipation. At large tip-sample distances, the AFM dissipation is sensitive to single electron charging, unlike STM which averages over a large number of such events. Thus, at AFM operation distances, the tunneling rate is far less to be detected as a tunneling signal by the STM. Similar to the case of quantum dots \cite{Stomp2005,Cockins2010} the amount of dissipated power is in the range of $meV/cycle$. It is worth to note that the effect might occur either from the tip side or sample side as confirmed by $dz/dV$ measurements (see Supplementary Figure 7).

\subsection{Conclusion}

Our low-temperature ($T=5\mathrm{K}$) AFM dissipation spectroscopy experiments showed multiple mechanical dissipation mechanisms over a topological insulator surface. The dissipation spectroscopy performed at tip-sample distances as large as several nm is sensitive to single electron tunneling into IPS. We attribute the observed dissipation peaks to charge fluctuation (van der Waals friction) in the system present when the IPS are occupied via single/few electron tunneling. The observation of IPS related dissipation features requires the suppression of Joule type of losses that is very small or absent on topologically protected surfaces due to lack of electron back-scattering. Joule and van der Waals type of energy losses are in the same order of magnitude on \BT surface. When an electron tunnels to the IPS, van der Waals dissipation increases due to increased charge fluctuations, while Joule dissipation decreases due to the screening effect. On the other hand, at larger magnetic fields (B$>$0.2T), we observed an increase in Joule dissipation due to the increase in electron back-scattering. As a result, dissipation peaks become less pronounced. The electronic characterization provided by the AFM mechanical dissipation peaks reported here may be used as an efficient and completely noninvasive tool for topological surface analysis, of considerable importance for nanotechnology. Finally, we demonstrated that pendulum AFM can address quantum effects in energy dissipation.

\section{Methods}

\subsection{Sample}
We used highly oriented \BT and single crystal \BT samples with resistivities: $\rho=0.1-5  $ $mOhm$ $\cdot cm$, and carrier mobilities: $\mu=3000 cm^2 \cdot V^{-1}s^{-1}$. Samples were cleaved under atmospheric conditions and immediately after transferred into the UHV chamber. Samples were heated to about $T=300^oC$ to remove water and weakly bounded molecules. After that, the crystal was introduced into the microscope chamber where it was cooled down to $T=5K$. We didn't observe significant differences between annealed and not annealed samples in our STM and pAFM measurements. Magnetic field experiments were done on single crystal \BT without magnetic impurities. We also used lump \BT flakes that may have more defects on the surface due to impurity doping.  

\subsection{Sensor}

All measurements were performed in ultra high vacuum ($p<10^{-10}\mathrm{mbar}$)  and at T=5K with metallic gold coated ATEC-non-contact cantilever from Nanosensors. Spring constant of the sensors was in the range of $\mathrm{k}=40-60\mathrm{N/m}$ and cantilevers were robust/stiff enough to use them as STM tips. The damping coefficient of such cantilevers was in the order of $\Gamma_0=10^{-10} \mathrm{kg/s}$ and normal forces were measured to be in the range of pN when we operate the sensor in STM mode.


\subsection{STM measurements}

STM measurements were done in constant current mode. During z-V spectroscopy of IPS on \BT(0001) the voltage was swept while the current feedback was active. Thus, the tip retracted from the surface as the voltage increased. We didn't use regular/ordinary STM tips that are metallic and rigid wires. The experiments were carried out with the flexible STM tip. We gain information about the forces by monitoring the static bending of these flexible probes. The scanning tip was metallic (gold coated) and free from uncompensated charges in order to perform proper STS measurements and to avoid static bending of the cantilever caused by electrostatic interaction. The same is valid to our AFM measurements, although AFM nominal working distance is further away from the surface as compared to STM.

\subsection{AFM measurements}

Force and dissipation measurements were done in pendulum-AFM mode (see Supplementary Information section S1 section for details) where cantilever oscillations were parallel to the measured surface. Thus, conservative and dissipative interactions can be measured in the non-contact regime. Lateral oscillation amplitude was kept constant by means of a Phase Locked Loop and vary from $A=200\mathrm{pm}$ to $-500\mathrm{pm}$. Damping coefficient as a measure of dissipation between tip and sample was measured at several tip-sample distances. The non-contact friction coefficient was calculated according to \cite{Cleveland1998}:
\begin{equation}
	\Gamma=\Gamma_0\left(\frac{A_\mathrm{exc}(z)}{A_{\mathrm{exc},0}}-\frac{f(z)}{f_0}\right),
\end{equation}
where $A_{exc}(z)$ and $f(z)$ are the distance-dependent excitation amplitude and resonance frequency of the cantilever, and the suffix 0 refers to the free cantilever. The distance $z=0$ corresponds to the point where the tip enters the contact regime, meaning that the cantilever driving signal is saturated and the tunneling current starts to rise. Friction coefficient can be converted into energy dissipation by: 
\begin{equation}
	P[eV/cycle]=\frac{2\pi^2A^2f_0}{e}\cdot \Gamma [kg/s],
\end{equation}
where $A$ is the oscillation amplitude and $e$ is the elementary charge.
  
\subsection{Simultaneous STM and AFM measurements}
We also performed simultaneous measurements where we used STM feedback and measured z-V by applying $A=30pm$ lateral oscillation to the tip.  At the very close distance of STM operation, much smaller amplitudes compared to regular AFM measurements have to be applied. So, the modulation currents due to the oscillation of the cantilever are negligible, and do not contribute to the STM feedback. In this mode, we gathered tunneling current and dissipation signals simultaneously as well as force information. 

\section{Acknowledgement}
We acknowledge fruitful discussions with Prof. Erio Tosatti. Basel group acknowledges financial support from the Swiss National Science Foundation (SNSF), the COST action Project MP1303, the SINERGIA Project CRSII2 136287/1, the European Union’s Horizon 2020 research and innovation program (ERC Advanced Grant, grant agreement No. 834402) and the Swiss Nanoscience Institute (Project No. P1301). O.G. acknowledges financial support from T\"{U}B\.{I}TAK project 114F036 and the COST action Project MP1303 (T\"{U}B\.{I}TAK112T818).

\section{Author contributions}
O. G. proposed the experiment. D. Y., M. K., and U. G. performed the experiments. E. M. coordinated the project. All authors discussed the results and contributed to the preparation of the paper.

%


\newpage

\newcommand{\beginsupplement}{%
	\setcounter{table}{0}
	\renewcommand{\thetable}{S\arabic{table}}%
	\setcounter{figure}{0}
	\renewcommand{\thefigure}{S\arabic{figure}}%
}

\section{Supplementary Information}
\beginsupplement

\subsection{S1. Short Description of Pendulum AFM and Dissipation Mechanisms }
Pendulum AFM (pAFM) is a unique home-built atomic force microscope dedicated to measure extremely small forces over surfaces by shearing the vacuum gap between tip and sample. To do so, extremely soft cantilevers ($k=10^{-5}$ - $10^{-3}$N/m) with high quality factors (Q $\approx$ $10^5$ - $10^6$ ) hover without contact perpendicularly to the sample surface in the pendulum geometry while avoiding snap into the contact (see Supplementary Figure 1(a) ). High Q-factor, together with extremely small k implies that the minimum detectable energy loss might be in the order of $\mu$eV/cycle - a value orders of magnitude smaller compared to standard AFM configurations. Pendulum AFM operates at 4.7K. It is also equipped with a scanning tunneling microscopy (STM) line, which allows the characterization of the electronic structure of the measured surface. The system is equipped with two UHV chambers: the preparation chamber to prepare atomically clean surfaces and the analysis chamber, where the microscope is located. The analysis chamber is equipped with perpendicular magnetic field spanning from B=-7T to +7T. 
In contrast to conventional contact friction AFM measurements, in pendulum geometry AFM, the tip and the sample are separated with a vacuum gap, and the end of the tip is oscillated above the surface and couples to the electronic or phononic type of excitations via non-contact interaction forces (see Supplementary Figure 1(b)). The non-invasive configuration and control over the tip distance and voltage allow for linear response theoretical descriptions and permits to distinguish experimentally between different channels of mechanical dissipation. Three main dissipation mechanisms that contribute are; (1) phononic and electronic friction whereby the moving tip drags the surface atomic and electronic chemical potential deformation and the energy is lost to the creation of phonons and (in metals) electron-hole pairs; (2) Joule dissipation from local currents induced when the charged tip oscillates over a resistive medium; and (3) van der Waals friction arises from the surface charge fluctuations.

\begin{figure}[!ht]\centering
\includegraphics[width=0.85\textwidth]{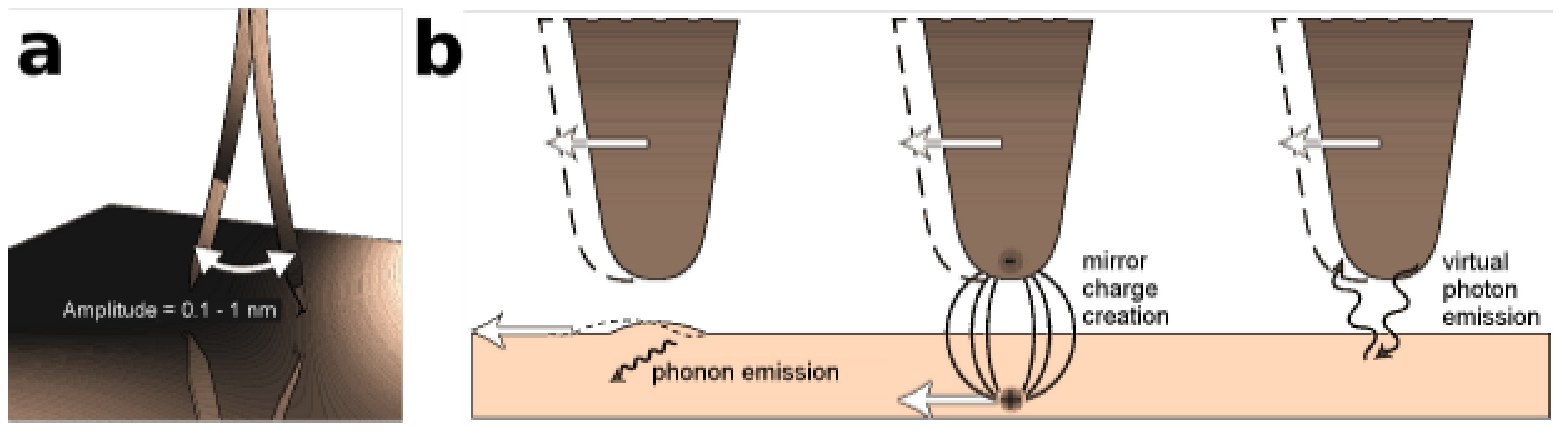}
\caption{\label{Notes1} 
\textbf{Short description of pendulum AFM.} a) Schematic drawing of pAFM cantilever oscillating in pendulum geometry over the sample surface. b)   Three main non-contact dissipation contributions are phononic and electronic friction whereby the surface atomic and electronic deformation is dragged by the moving tip and the energy is lost to the creation of phonons and (in metals) electron-hole pairs; Joule dissipation from local currents induced when the charged tip oscillates over a resistive medium; van der Waals friction arises from the surface charge fluctuations. }
\end{figure}

\begin{figure}[!ht]\centering
\includegraphics[width=0.8\textwidth]{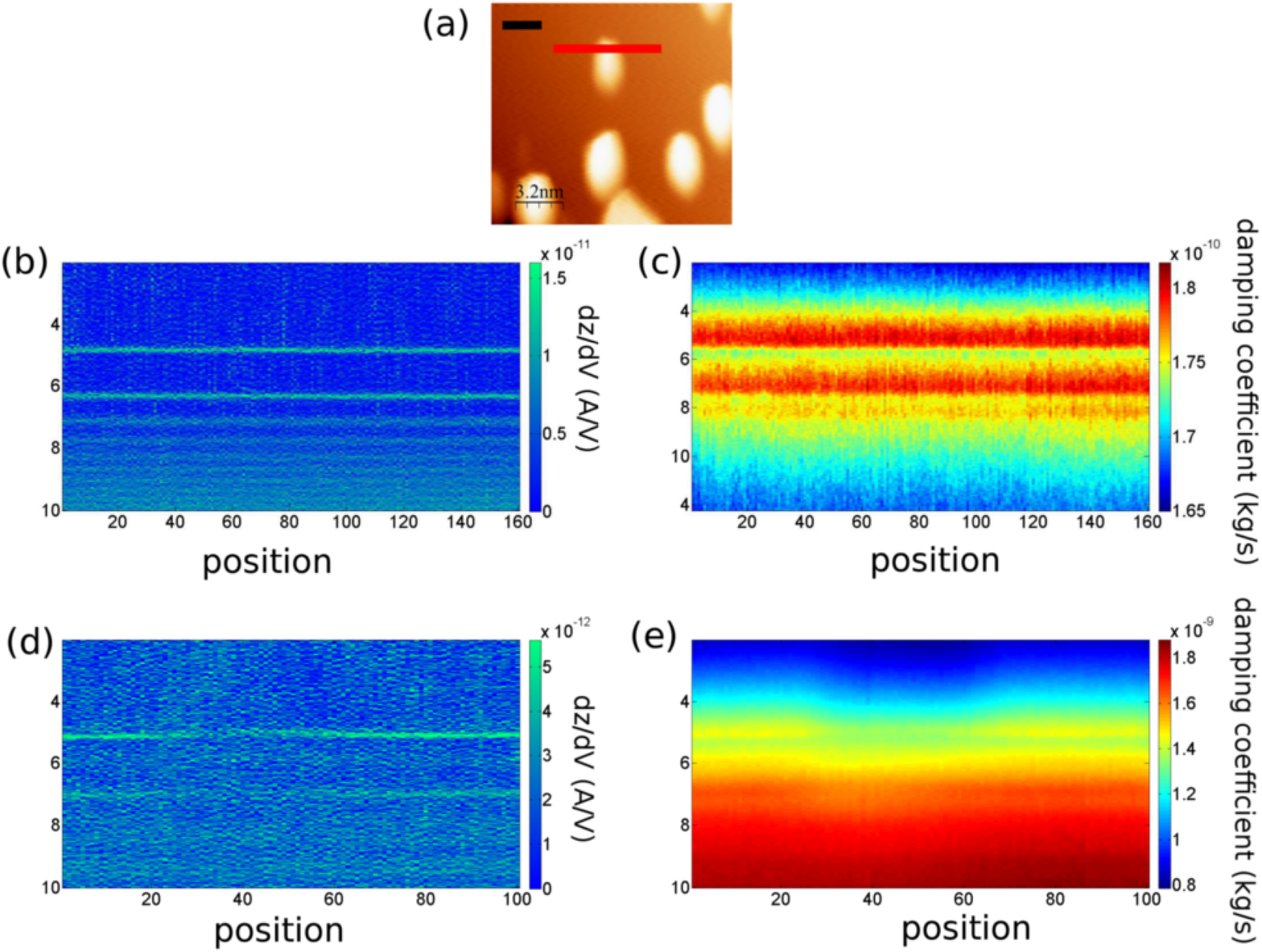}
\caption{\label{Notes2} 
\textbf{Position dependent $dz/dV$ and dissipation spectroscopy performed on flat (b), (c) and defected (d),(e) area with an oscillating tip.} (a) STM data of \BT surface shows structural defects. Spectroscopies were performed on a line on clean (black line) and structurally defected area (red line) to see the spatial dependence of dz/dV and dissipation curves. (b) and (c) shows that dissipation signal is correlated with dz/dV data and both of them don't show position dependent change. (d) and (e) shows that energy of the first IPS changes slightly on the defect, dissipation signal shows more visible.  Tunneling parameters, $I_t= 95 pA , V_s=2 V $. Oscillation amplitude $A_{osc}=30 pm$.
}
\end{figure}

\begin{figure}[!ht]\centering
\includegraphics[width=0.65\textwidth]{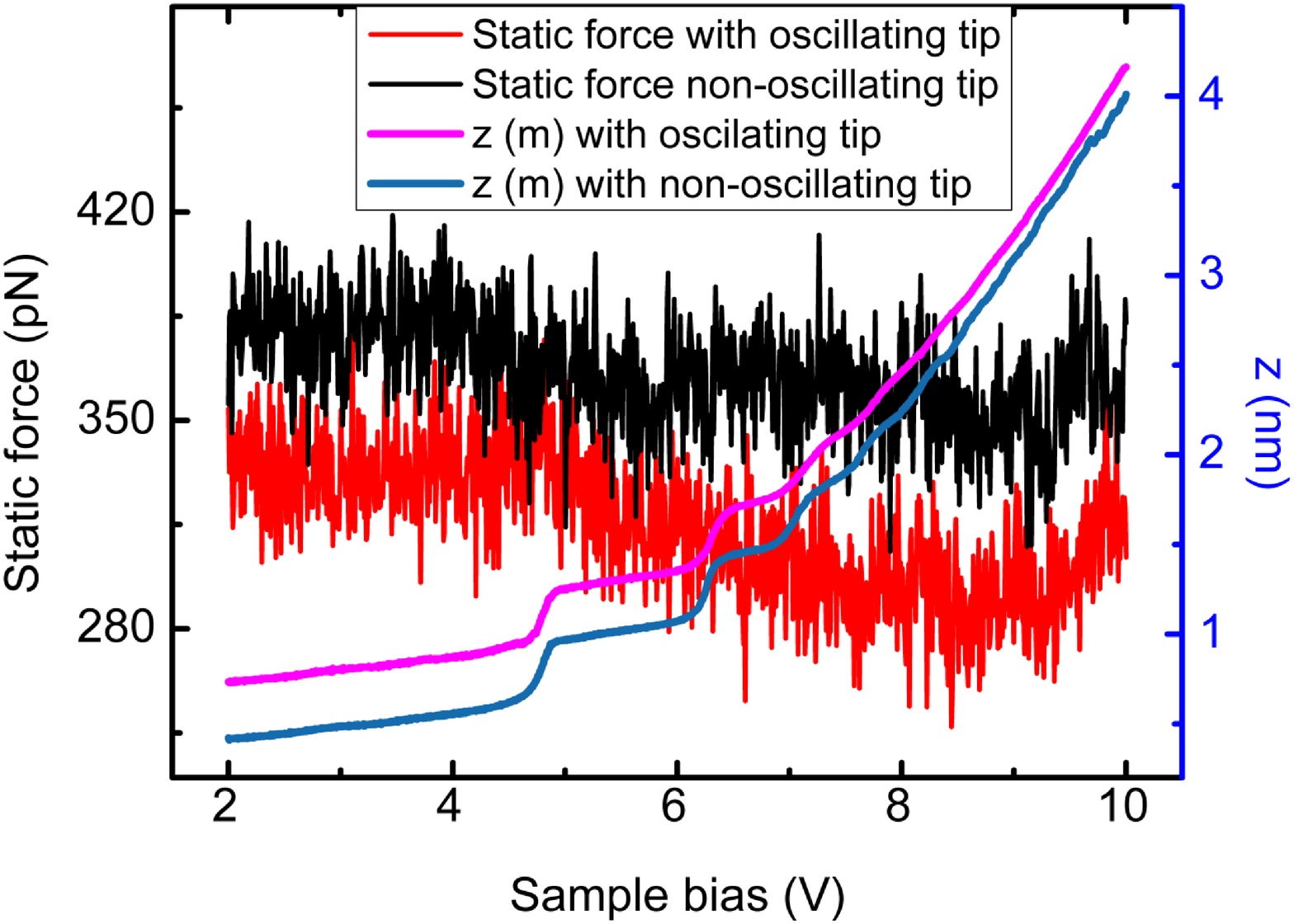}
\caption{\label{Notes3} 
\textbf{Static forces during STM measurements.} Static force is measured simultaneously during z-V spectroscopy. Static force is in the range of $10^{-10} N $ for the measurements with oscillating (red) and non-oscillating (black) tip. z-V curves show that tip retracts $\sim 0.3$  nm when the tip is oscillated with lateral oscillation amplitude 30 pm. Static force slightly decreases due to the retraction of an oscillated tip. Tunneling parameters; $I_t= 96 pA$ and $V_s=2V$ for both measurements. 
}
\end{figure}

\begin{figure}[!ht]\centering
\includegraphics[width=0.65\textwidth]{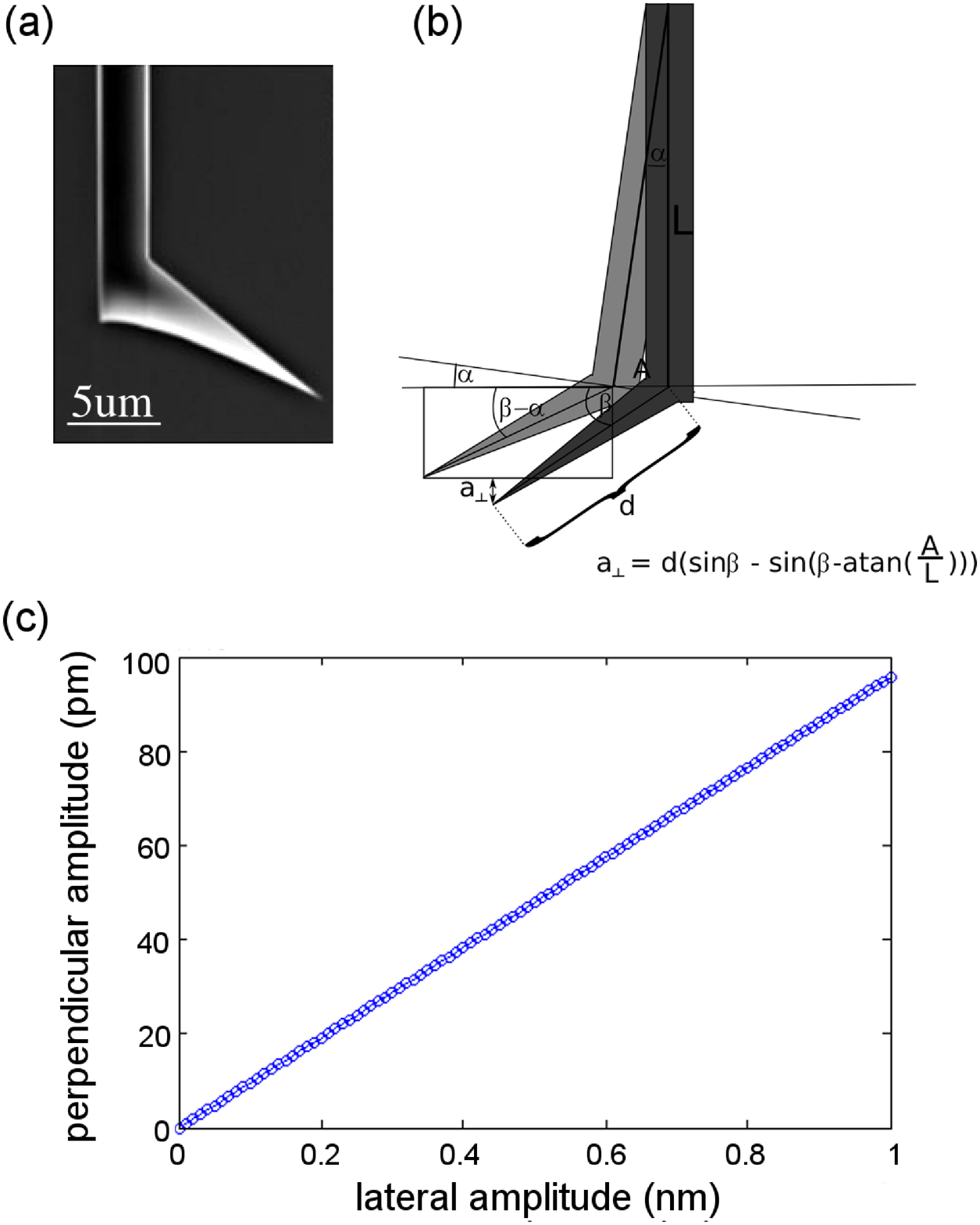}
\caption{\label{Notes3} 
\textbf{An asymmetric pendulum AFM tip.} (a) SEM micrograph of the Nanosensors - ATEC NcAu cantilever tip. (b) lateral oscillation with amplitude A leads to non-negligible perpendicular amplitude of oscillation $a_{\perp}$ (L is the cantilever length). (c) $a_{\perp}$ versus A for ATEC NcAu cantilever.}
\end{figure}

\begin{figure}[!ht]\centering
\includegraphics[width=0.85\textwidth]{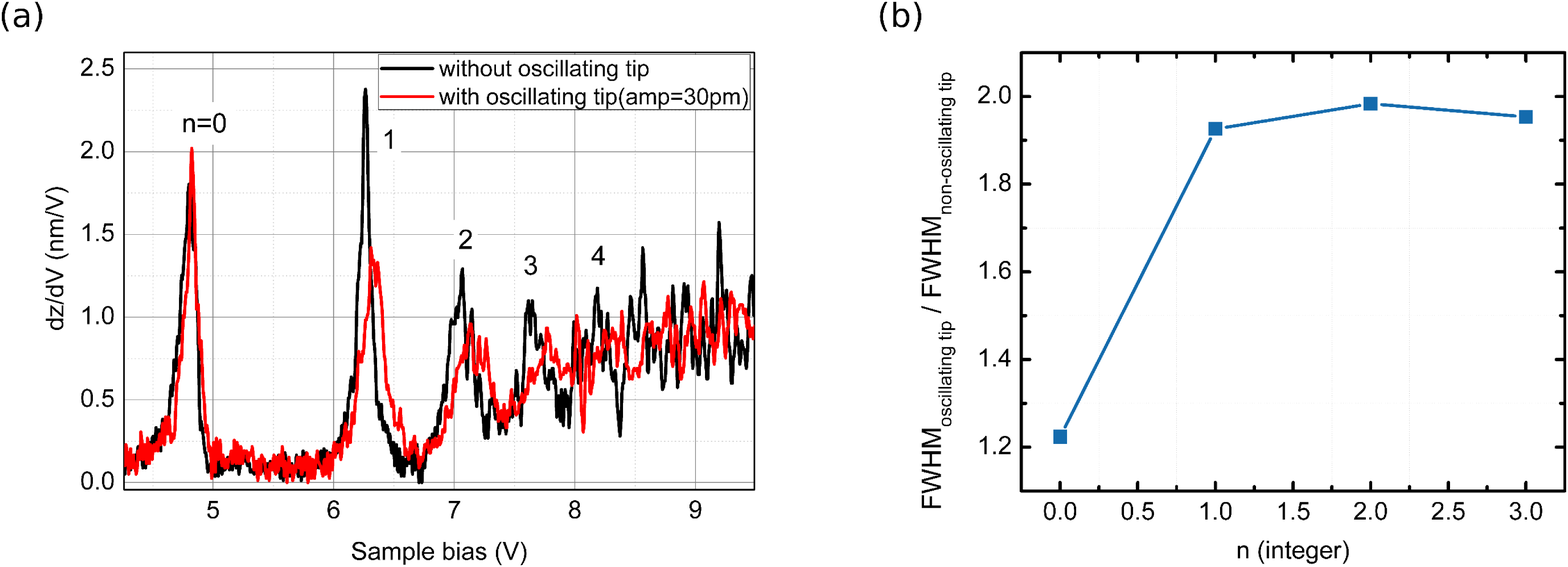}
\caption{\label{Notes4} 
\textbf{dz/dV(V) spectroscopy showing image potential states measured by static and oscillating STM tip (a).} On (b) the increase of FWHM of IPS peaks in dz/dV(V) spectra are shown for  dynamic situation compared to the static case. FWHM of the IPS peaks are measured from the dz/dV(V) data and the states numbered with n=1,2,3 get broadened by a factor of two, whereas state n=0 is almost unaffected by tip oscillations. 
}
\end{figure}

\clearpage

\subsection{S2. Displacement current and Joule dissipation:}

Dissipated power
\begin{equation}
P=I^2 \cdot R = V_B \cdot I
\end{equation}
where $I, R, V_B$ stand for displacement current, resistance and bias voltage. When a voltage $V_B$ is applied between tip and sample the displacement current is equal to:

\begin{equation}
i_d = \dot q(t)=\dot C V_B = \frac{\partial C}{\partial z} \frac{\partial z}{\partial t} V_B
\end{equation}

where C is tip-sample capacitance and the tip oscillations $z(t) = A sin (\omega t + \phi)$. Thus:

\begin{equation}
P = R \cdot i_d^2 = R \left(\frac{\partial C}{\partial z}\right)^2 V_B^2 \left(\frac{\partial z}{\partial t}\right)^2
\end{equation}

where $\gamma = R \left(\frac{\partial C}{\partial z}\right)^2 V_B^2$ is the effective damping coefficient proportional to the resistance of the sample. The dissipated power:
\begin{equation}
P=R \left(\frac{\partial C}{\partial z}\right)^2 V_B^2 \left(-A \omega^2 sin(\omega t + \phi)\right)
\end{equation}
and dissipated power averaged over one period of oscillations:
\begin{equation}
<P> = \frac{1}{2} R A^2 \omega^2 V_B^2 \left(\frac{\partial C}{\partial z}\right)^2
\end{equation}

\clearpage

\subsection{S3. Frequency shift ($\Delta f$) and two capacitor model:}

\begin{figure}[!ht]\centering
\includegraphics[width=0.85\textwidth]{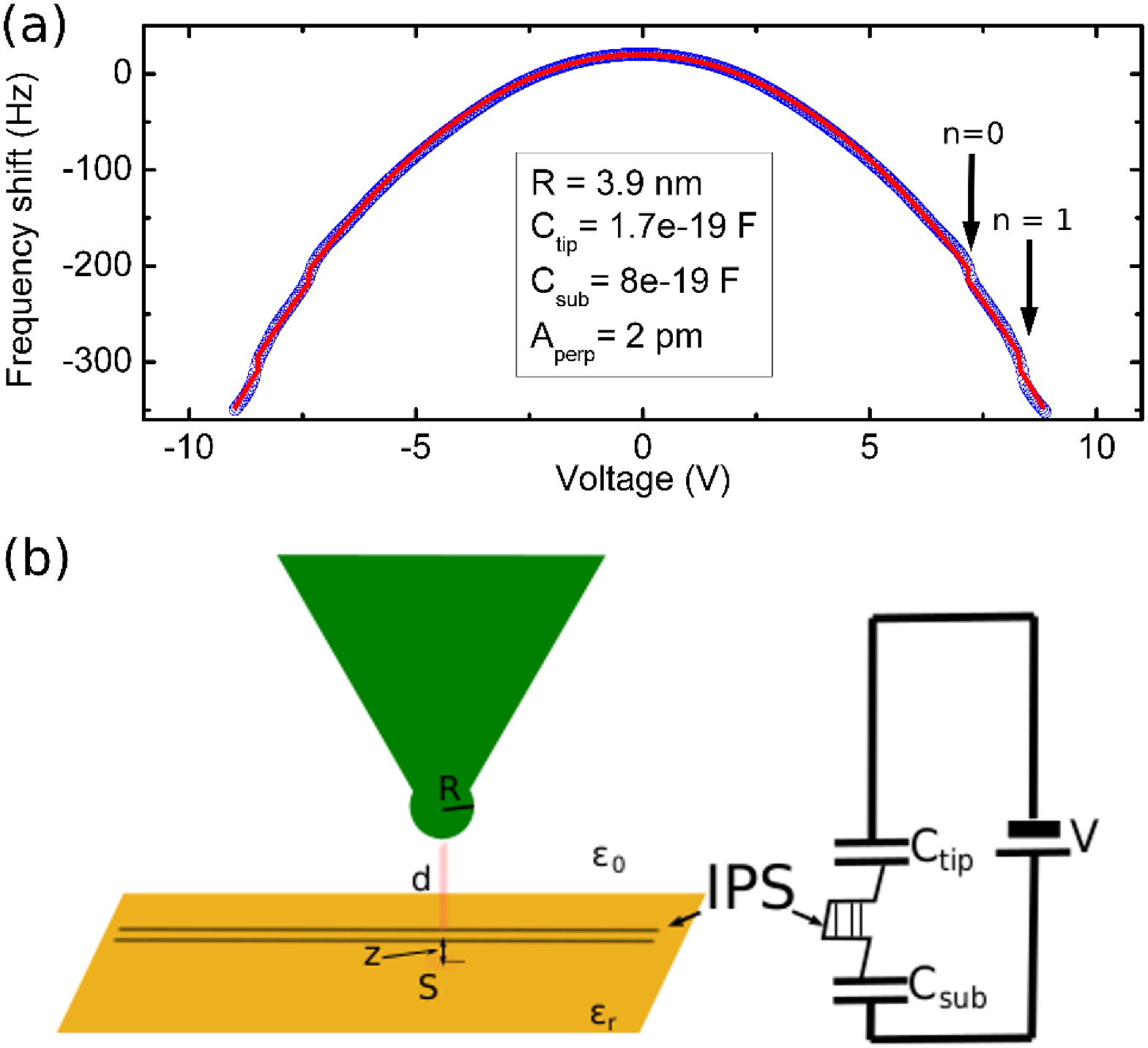}
\caption{\label{Notes5} 
\textbf{Voltage dependent frequency shift spectra at $d=5nm$ distance over the surface.}  On (a) frequency shift parabola shows some singularities at the biases when single electron charging of IPS occurs. The blue points are experimental data, while the red line is simulated $\Delta f$ fit. The fit allows to estimate the tip-IPS ($C_{tip}$) and IPS-sample ($C_{sub}$) capacitances. The schmatics of the tip-sample geometry and  the equvalent electrical circuit is shown in b, where $R$, $d$, $z$ and $S$ stands for tip radius, tip distance to the IPS, distance of IPS to the surface and tip projected "active" surface area, respectively. $\epsilon_0$ and $\epsilon_r=75$ are vacuum permittivity and dilelectric constant of \BT, respectively.}
\end{figure}

The $C_{tip} = \frac{2\pi R^2}{d}=0.17aF$ is estimated using spherical tip over the plane. The tip radius was fixed to $R=3.9nm$, in agreement with dissipation map shown in Figure 4(b). \\
The force is calculated according to formula \cite{Stomp2005};  \\

$F=\frac{1}{(C_{tip}+C_{sub})^2}\frac{\partial C_{tip}}{\partial z} \left(\frac{q^2}{2}-C_{sub}q (V-V_{CPD})+\frac{1}{2}C_{sub}^2(V-V_{CPD})^2\right)$\\
where $q=ne$. \\
Next the frequency shift was calculated \cite{Stomp2005} and fit to the experimental data:\\

$\Delta f(d) = \frac{f_0^2}{kA} \int_{0}^{1/f_0} F[d+Acos(2 \pi f_0 t)]cos(2 \pi f_0 t) dt  $  \\

The fitting parameters were: IPS-sample capacitance $C_{sub}=0.8aF$ and perpendicular component of oscillation amplitude $A=2pm$. Thus, $\frac{C_{tip}}{C_{sub}} = 0.2$

In order to validate the model, we estimate the distance $z$, namely distance above the surface were IPS are located. Plane capacitor model was assumed: 
$C_{sub}=\frac{\epsilon_r \epsilon_0 S}{z}$, where $\epsilon_r=\epsilon(0)=75$ for \BT was taken after [W. Richter, H. Koehler, C.R. Becker, A Raman and infrared investigation of phonons in the rhombohedral V2-VI3 compounds, Phys. Status Solidi (b) 84 (1977) 619]. "Active" surface area of interaction was assumed to be equal $S=\pi z^2$. \\
Taking everything into consideration, we get: $z=\frac{2R^2}{0.2d\epsilon_r}=0.4 nm$, which is very realistic estimate.

\clearpage

\subsection{S4. Measuring IPS of the tip with z-V and dissipation spectroscopy }
In our system, the sample is biased, and the tip is grounded. When the sample is biased positively, IPS of the sample is measured with z-V spectroscopy. IPS of the tip is probed if the sample is negatively biased (see Supplementary Figure 7). The first peak of the dz/dV (V) data is related to the work function of the sample, and its energy may depend on the tip material and shape. A slight difference between the energy of the first peak of IPS of the tip and the sample can be seen in Supplementary Figure 7. 

Work function difference between the tip and the sample can be measured as contact potential difference (CPD) using AFM. CPD value between the tip and the sample can be found by measuring the bias where the dissipation spectrum is minimum. If the tip and the sample are made from the same material, the CPD value would be 0 V.  Supplementary Figure 8. shows three different energy dissipation spectra with different CPD values. The dissipation spectra are symmetric if CPD is 0V and asymmetry is measured to be increasing with an increased CPD.

\begin{figure}[!ht]\centering
\includegraphics[width=0.8\textwidth]{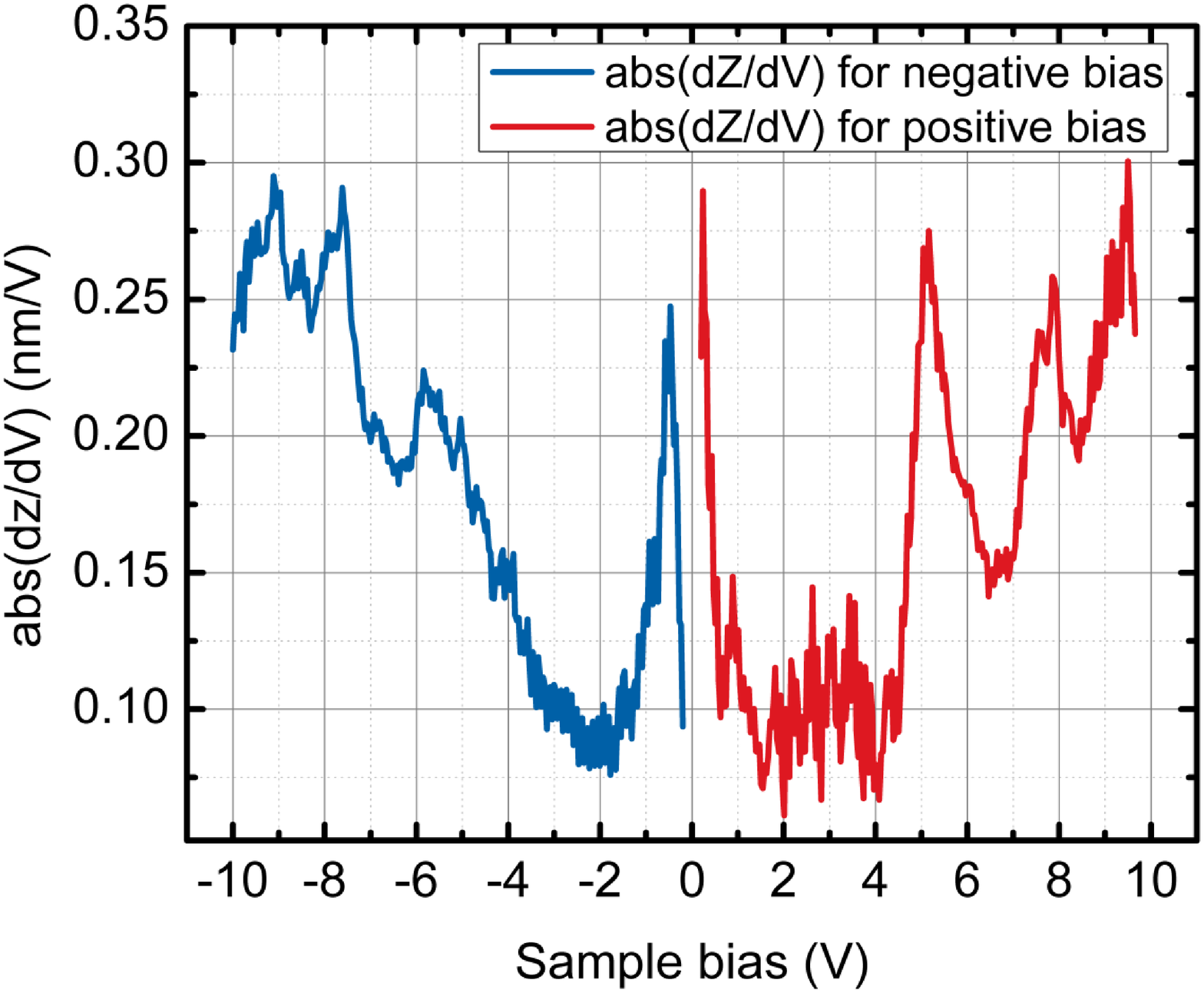}
\caption{\label{Notes6_a } dz/dV(V) for positive and negative sample bias. The curve shows IPS of the sample at positive biases and IPS of the tip at negative biases. The spectra showing IPS of the tip doesn't have very sharp peaks; however, it still shows that the energies of the states are comparable to the energies of the IPS of the sample. Slight asymmetry in the energies of the first peaks is visible. The energy of the first peak of the IPS of the tip is slightly higher than the energy of the first peak of the IPS of the sample.
}
\end{figure}

\begin{figure}[!ht]\centering
\includegraphics[width=0.85\textwidth]{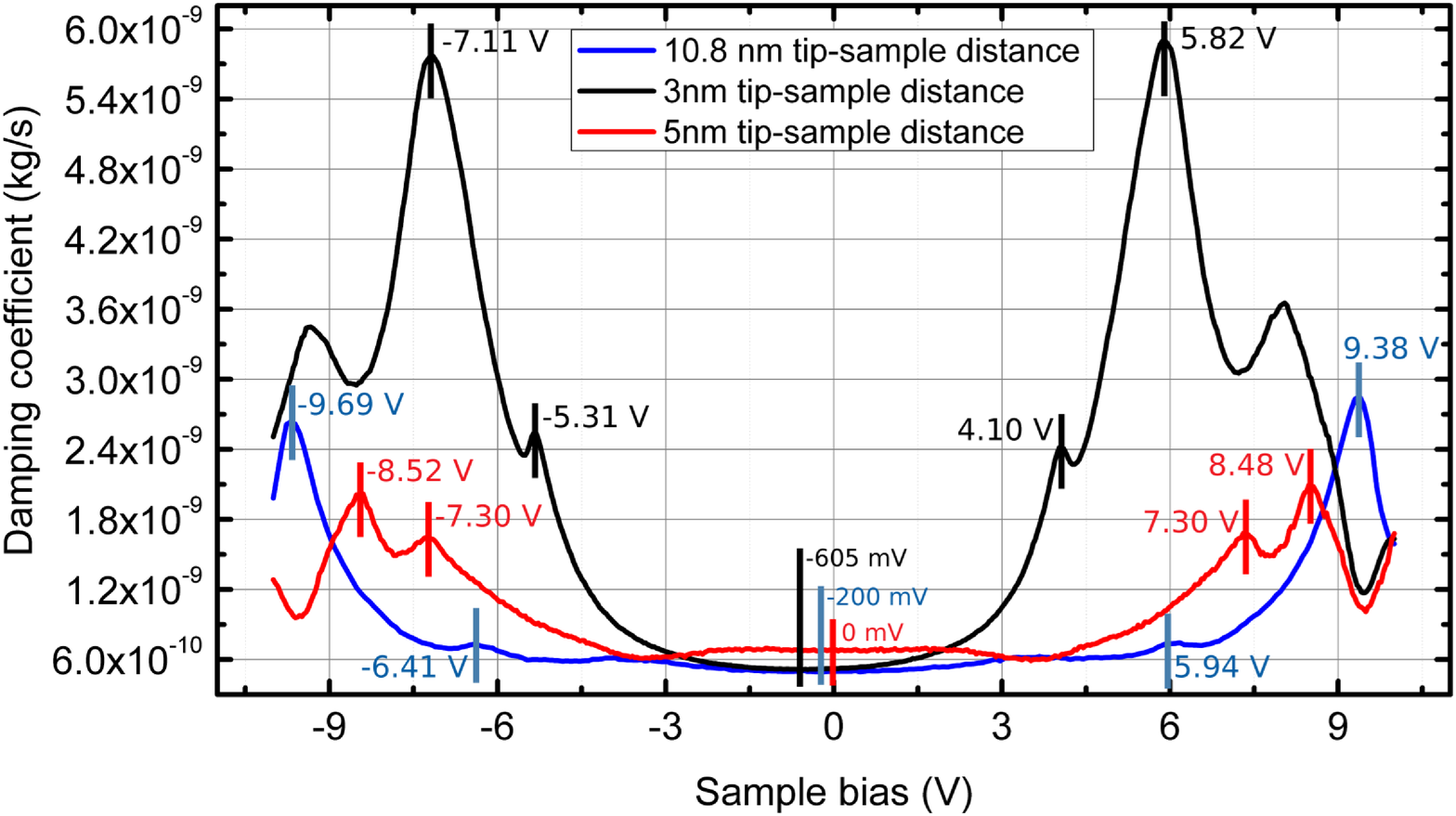}
\caption{\label{Notes6_b} 
\textbf{Energy  dissipation versus bias voltage.} Three dissipation curves are plotted in one graph to show the symmetric and asymmetric cases around 0V. Asymmetry is proportional to the measured CPD values. Measured CPD value may vary depending on the work functions of the tip and the sample. The work function of bulk Au, Cr, and crystalline \BT are 5.1, 4.5, and 5.3 eV, respectively. Slight asymmetry is seen in a blue curve, where CPD is measured -200mV. The asymmetry is visible on the black curve where CPD is measured to be -605mV. Cr/Au mixture can lead to a work function value between the work function of pure Au and Cr. The blue curve shows that CPD value is equal to the work function difference between pure Au and \BT. \BT flake can be picked up by the tip during the measurements, and this results with symmetric dissipation signal as shown in the red curve. When the sample and tip are made of the same material, CPD value is measured to be 0V. 
}
\end{figure}


\begin{thebibliography}{39}%
\makeatletter
\providecommand \@ifxundefined [1]{%
 \@ifx{#1\undefined}
}%
\providecommand \@ifnum [1]{%
 \ifnum #1\expandafter \@firstoftwo
 \else \expandafter \@secondoftwo
 \fi
}%
\providecommand \@ifx [1]{%
 \ifx #1\expandafter \@firstoftwo
 \else \expandafter \@secondoftwo
 \fi
}%
\providecommand \natexlab [1]{#1}%
\providecommand \enquote  [1]{``#1''}%
\providecommand \bibnamefont  [1]{#1}%
\providecommand \bibfnamefont [1]{#1}%
\providecommand \citenamefont [1]{#1}%
\providecommand \href@noop [0]{\@secondoftwo}%
\providecommand \href [0]{\begingroup \@sanitize@url \@href}%
\providecommand \@href[1]{\@@startlink{#1}\@@href}%
\providecommand \@@href[1]{\endgroup#1\@@endlink}%
\providecommand \@sanitize@url [0]{\catcode `\\12\catcode `\$12\catcode
  `\&12\catcode `\#12\catcode `\^12\catcode `\_12\catcode `\%12\relax}%
\providecommand \@@startlink[1]{}%
\providecommand \@@endlink[0]{}%
\providecommand \url  [0]{\begingroup\@sanitize@url \@url }%
\providecommand \@url [1]{\endgroup\@href {#1}{\urlprefix }}%
\providecommand \urlprefix  [0]{URL }%
\providecommand \Eprint [0]{\href }%
\providecommand \doibase [0]{http://dx.doi.org/}%
\providecommand \selectlanguage [0]{\@gobble}%
\providecommand \bibinfo  [0]{\@secondoftwo}%
\providecommand \bibfield  [0]{\@secondoftwo}%
\providecommand \translation [1]{[#1]}%
\providecommand \BibitemOpen [0]{}%
\providecommand \bibitemStop [0]{}%
\providecommand \bibitemNoStop [0]{.\EOS\space}%
\providecommand \EOS [0]{\spacefactor3000\relax}%
\providecommand \BibitemShut  [1]{\csname bibitem#1\endcsname}%
\let\auto@bib@innerbib\@empty
\bibitem [{\citenamefont {Chen}\ \emph {et~al.}(2009)\citenamefont {Chen},
  \citenamefont {Analytis}, \citenamefont {Chu}, \citenamefont {Liu},
  \citenamefont {Mo}, \citenamefont {Qi}, \citenamefont {Zhang}, \citenamefont
  {Lu}, \citenamefont {Dai}, \citenamefont {Fang}, \citenamefont {Zhang},
  \citenamefont {Fisher}, \citenamefont {Hussain},\ and\ \citenamefont
  {Shen}}]{Chen2009}%
  \BibitemOpen
  \bibfield  {author} {\bibinfo {author} {\bibfnamefont {Y.~L.}\ \bibnamefont
  {Chen}}, \bibinfo {author} {\bibfnamefont {J.~G.}\ \bibnamefont {Analytis}},
  \bibinfo {author} {\bibfnamefont {J.-H.}\ \bibnamefont {Chu}}, \bibinfo
  {author} {\bibfnamefont {Z.~K.}\ \bibnamefont {Liu}}, \bibinfo {author}
  {\bibfnamefont {S.-K.}\ \bibnamefont {Mo}}, \bibinfo {author} {\bibfnamefont
  {X.~L.}\ \bibnamefont {Qi}}, \bibinfo {author} {\bibfnamefont {H.~J.}\
  \bibnamefont {Zhang}}, \bibinfo {author} {\bibfnamefont {D.~H.}\ \bibnamefont
  {Lu}}, \bibinfo {author} {\bibfnamefont {X.}~\bibnamefont {Dai}}, \bibinfo
  {author} {\bibfnamefont {Z.}~\bibnamefont {Fang}}, \bibinfo {author}
  {\bibfnamefont {S.~C.}\ \bibnamefont {Zhang}}, \bibinfo {author}
  {\bibfnamefont {I.~R.}\ \bibnamefont {Fisher}}, \bibinfo {author}
  {\bibfnamefont {Z.}~\bibnamefont {Hussain}}, \ and\ \bibinfo {author}
  {\bibfnamefont {Z.-X.}\ \bibnamefont {Shen}},\ }\href {\doibase
  10.1126/science.1173034} {\bibfield  {journal} {\bibinfo  {journal}
  {Science}\ }\textbf {\bibinfo {volume} {325}},\ \bibinfo {pages} {178}
  (\bibinfo {year} {2009})}\BibitemShut {NoStop}%
\bibitem [{\citenamefont {Hasan}\ and\ \citenamefont {Kane}(2010)}]{Hasan2010}%
  \BibitemOpen
  \bibfield  {author} {\bibinfo {author} {\bibfnamefont {M.~Z.}\ \bibnamefont
  {Hasan}}\ and\ \bibinfo {author} {\bibfnamefont {C.~L.}\ \bibnamefont
  {Kane}},\ }\href {\doibase 10.1103/RevModPhys.82.3045} {\bibfield  {journal}
  {\bibinfo  {journal} {Rev. Mod. Phys.}\ }\textbf {\bibinfo {volume} {82}},\
  \bibinfo {pages} {3045} (\bibinfo {year} {2010})}\BibitemShut {NoStop}%
\bibitem [{\citenamefont {Seo}\ \emph {et~al.}(2010)\citenamefont {Seo},
  \citenamefont {Roushan}, \citenamefont {Beidenkopf}, \citenamefont {Hor},
  \citenamefont {Cava},\ and\ \citenamefont {Yazdani}}]{Seo2010}%
  \BibitemOpen
  \bibfield  {author} {\bibinfo {author} {\bibfnamefont {J.}~\bibnamefont
  {Seo}}, \bibinfo {author} {\bibfnamefont {P.}~\bibnamefont {Roushan}},
  \bibinfo {author} {\bibfnamefont {H.}~\bibnamefont {Beidenkopf}}, \bibinfo
  {author} {\bibfnamefont {Y.~S.}\ \bibnamefont {Hor}}, \bibinfo {author}
  {\bibfnamefont {R.~J.}\ \bibnamefont {Cava}}, \ and\ \bibinfo {author}
  {\bibfnamefont {A.}~\bibnamefont {Yazdani}},\ }\href
  {http://dx.doi.org/10.1038/nature09189} {\bibfield  {journal} {\bibinfo
  {journal} {Nature}\ }\textbf {\bibinfo {volume} {466}},\ \bibinfo {pages}
  {343 EP } (\bibinfo {year} {2010})}\BibitemShut {NoStop}%
\bibitem [{\citenamefont {Zhang}\ \emph {et~al.}(2009)\citenamefont {Zhang},
  \citenamefont {Cheng}, \citenamefont {Chen}, \citenamefont {Jia},
  \citenamefont {Ma}, \citenamefont {He}, \citenamefont {Wang}, \citenamefont
  {Zhang}, \citenamefont {Dai}, \citenamefont {Fang}, \citenamefont {Xie},\
  and\ \citenamefont {Xue}}]{Zhang2009B}%
  \BibitemOpen
  \bibfield  {author} {\bibinfo {author} {\bibfnamefont {T.}~\bibnamefont
  {Zhang}}, \bibinfo {author} {\bibfnamefont {P.}~\bibnamefont {Cheng}},
  \bibinfo {author} {\bibfnamefont {X.}~\bibnamefont {Chen}}, \bibinfo {author}
  {\bibfnamefont {J.-F.}\ \bibnamefont {Jia}}, \bibinfo {author} {\bibfnamefont
  {X.}~\bibnamefont {Ma}}, \bibinfo {author} {\bibfnamefont {K.}~\bibnamefont
  {He}}, \bibinfo {author} {\bibfnamefont {L.}~\bibnamefont {Wang}}, \bibinfo
  {author} {\bibfnamefont {H.}~\bibnamefont {Zhang}}, \bibinfo {author}
  {\bibfnamefont {X.}~\bibnamefont {Dai}}, \bibinfo {author} {\bibfnamefont
  {Z.}~\bibnamefont {Fang}}, \bibinfo {author} {\bibfnamefont {X.}~\bibnamefont
  {Xie}}, \ and\ \bibinfo {author} {\bibfnamefont {Q.-K.}\ \bibnamefont
  {Xue}},\ }\href {\doibase 10.1103/PhysRevLett.103.266803} {\bibfield
  {journal} {\bibinfo  {journal} {Phys. Rev. Lett.}\ }\textbf {\bibinfo
  {volume} {103}},\ \bibinfo {pages} {266803} (\bibinfo {year}
  {2009})}\BibitemShut {NoStop}%
\bibitem [{\citenamefont {He}\ \emph {et~al.}(2011)\citenamefont {He},
  \citenamefont {Wang}, \citenamefont {Zhang}, \citenamefont {Sou},
  \citenamefont {Wong}, \citenamefont {Wang}, \citenamefont {Lu}, \citenamefont
  {Shen},\ and\ \citenamefont {Zhang}}]{Hong2011}%
  \BibitemOpen
  \bibfield  {author} {\bibinfo {author} {\bibfnamefont {H.-T.}\ \bibnamefont
  {He}}, \bibinfo {author} {\bibfnamefont {G.}~\bibnamefont {Wang}}, \bibinfo
  {author} {\bibfnamefont {T.}~\bibnamefont {Zhang}}, \bibinfo {author}
  {\bibfnamefont {I.-K.}\ \bibnamefont {Sou}}, \bibinfo {author} {\bibfnamefont
  {G.~K.~L.}\ \bibnamefont {Wong}}, \bibinfo {author} {\bibfnamefont {J.-N.}\
  \bibnamefont {Wang}}, \bibinfo {author} {\bibfnamefont {H.-Z.}\ \bibnamefont
  {Lu}}, \bibinfo {author} {\bibfnamefont {S.-Q.}\ \bibnamefont {Shen}}, \ and\
  \bibinfo {author} {\bibfnamefont {F.-C.}\ \bibnamefont {Zhang}},\ }\href
  {\doibase 10.1103/PhysRevLett.106.166805} {\bibfield  {journal} {\bibinfo
  {journal} {Phys. Rev. Lett.}\ }\textbf {\bibinfo {volume} {106}},\ \bibinfo
  {pages} {166805} (\bibinfo {year} {2011})}\BibitemShut {NoStop}%
\bibitem [{\citenamefont {Straub}\ and\ \citenamefont
  {Himpsel}(1986)}]{Himpsel1986}%
  \BibitemOpen
  \bibfield  {author} {\bibinfo {author} {\bibfnamefont {D.}~\bibnamefont
  {Straub}}\ and\ \bibinfo {author} {\bibfnamefont {F.~J.}\ \bibnamefont
  {Himpsel}},\ }\href {\doibase 10.1103/PhysRevB.33.2256} {\bibfield  {journal}
  {\bibinfo  {journal} {Phys. Rev. B}\ }\textbf {\bibinfo {volume} {33}},\
  \bibinfo {pages} {2256} (\bibinfo {year} {1986})}\BibitemShut {NoStop}%
\bibitem [{\citenamefont {Dose}(1987)}]{Dose1987}%
  \BibitemOpen
  \bibfield  {author} {\bibinfo {author} {\bibfnamefont {V.}~\bibnamefont
  {Dose}},\ }\href {http://stacks.iop.org/1402-4896/36/i=4/a=009} {\bibfield
  {journal} {\bibinfo  {journal} {Physica Scripta}\ }\textbf {\bibinfo {volume}
  {36}},\ \bibinfo {pages} {669} (\bibinfo {year} {1987})}\BibitemShut
  {NoStop}%
\bibitem [{\citenamefont {Echenique}\ and\ \citenamefont
  {Uranga}(1991)}]{Echenique1991}%
  \BibitemOpen
  \bibfield  {author} {\bibinfo {author} {\bibfnamefont {P.}~\bibnamefont
  {Echenique}}\ and\ \bibinfo {author} {\bibfnamefont {M.}~\bibnamefont
  {Uranga}},\ }\href {\doibase https://doi.org/10.1016/0039-6028(91)90118-C}
  {\bibfield  {journal} {\bibinfo  {journal} {Surface Science}\ }\textbf
  {\bibinfo {volume} {247}},\ \bibinfo {pages} {125 } (\bibinfo {year}
  {1991})}\BibitemShut {NoStop}%
\bibitem [{\citenamefont {Berthold}\ \emph {et~al.}(2002)\citenamefont
  {Berthold}, \citenamefont {H\"ofer}, \citenamefont {Feulner}, \citenamefont
  {Chulkov}, \citenamefont {Silkin},\ and\ \citenamefont
  {Echenique}}]{Berthold2002}%
  \BibitemOpen
  \bibfield  {author} {\bibinfo {author} {\bibfnamefont {W.}~\bibnamefont
  {Berthold}}, \bibinfo {author} {\bibfnamefont {U.}~\bibnamefont {H\"ofer}},
  \bibinfo {author} {\bibfnamefont {P.}~\bibnamefont {Feulner}}, \bibinfo
  {author} {\bibfnamefont {E.~V.}\ \bibnamefont {Chulkov}}, \bibinfo {author}
  {\bibfnamefont {V.~M.}\ \bibnamefont {Silkin}}, \ and\ \bibinfo {author}
  {\bibfnamefont {P.~M.}\ \bibnamefont {Echenique}},\ }\href {\doibase
  10.1103/PhysRevLett.88.056805} {\bibfield  {journal} {\bibinfo  {journal}
  {Phys. Rev. Lett.}\ }\textbf {\bibinfo {volume} {88}},\ \bibinfo {pages}
  {056805} (\bibinfo {year} {2002})}\BibitemShut {NoStop}%
\bibitem [{\citenamefont {Wahl}\ \emph {et~al.}(2003)\citenamefont {Wahl},
  \citenamefont {Schneider}, \citenamefont {Diekh\"oner}, \citenamefont
  {Vogelgesang},\ and\ \citenamefont {Kern}}]{Wahl2003}%
  \BibitemOpen
  \bibfield  {author} {\bibinfo {author} {\bibfnamefont {P.}~\bibnamefont
  {Wahl}}, \bibinfo {author} {\bibfnamefont {M.~A.}\ \bibnamefont {Schneider}},
  \bibinfo {author} {\bibfnamefont {L.}~\bibnamefont {Diekh\"oner}}, \bibinfo
  {author} {\bibfnamefont {R.}~\bibnamefont {Vogelgesang}}, \ and\ \bibinfo
  {author} {\bibfnamefont {K.}~\bibnamefont {Kern}},\ }\href {\doibase
  10.1103/PhysRevLett.91.106802} {\bibfield  {journal} {\bibinfo  {journal}
  {Phys. Rev. Lett.}\ }\textbf {\bibinfo {volume} {91}},\ \bibinfo {pages}
  {106802} (\bibinfo {year} {2003})}\BibitemShut {NoStop}%
\bibitem [{\citenamefont {Schouteden}\ and\ \citenamefont
  {Van~Haesendonck}(2009)}]{Schouteden2009}%
  \BibitemOpen
  \bibfield  {author} {\bibinfo {author} {\bibfnamefont {K.}~\bibnamefont
  {Schouteden}}\ and\ \bibinfo {author} {\bibfnamefont {C.}~\bibnamefont
  {Van~Haesendonck}},\ }\href {\doibase 10.1103/PhysRevLett.103.266805}
  {\bibfield  {journal} {\bibinfo  {journal} {Phys. Rev. Lett.}\ }\textbf
  {\bibinfo {volume} {103}},\ \bibinfo {pages} {266805} (\bibinfo {year}
  {2009})}\BibitemShut {NoStop}%
\bibitem [{\citenamefont {Niesner}\ and\ \citenamefont
  {Fauster}(2014)}]{Niesner2014C}%
  \BibitemOpen
  \bibfield  {author} {\bibinfo {author} {\bibfnamefont {D.}~\bibnamefont
  {Niesner}}\ and\ \bibinfo {author} {\bibfnamefont {T.}~\bibnamefont
  {Fauster}},\ }\href {http://stacks.iop.org/0953-8984/26/i=39/a=393001}
  {\bibfield  {journal} {\bibinfo  {journal} {Journal of Physics: Condensed
  Matter}\ }\textbf {\bibinfo {volume} {26}},\ \bibinfo {pages} {393001}
  (\bibinfo {year} {2014})}\BibitemShut {NoStop}%
\bibitem [{\citenamefont {Sobota}\ \emph {et~al.}(2012)\citenamefont {Sobota},
  \citenamefont {Yang}, \citenamefont {Analytis}, \citenamefont {Chen},
  \citenamefont {Fisher}, \citenamefont {Kirchmann},\ and\ \citenamefont
  {Shen}}]{Sobota2012}%
  \BibitemOpen
  \bibfield  {author} {\bibinfo {author} {\bibfnamefont {J.~A.}\ \bibnamefont
  {Sobota}}, \bibinfo {author} {\bibfnamefont {S.}~\bibnamefont {Yang}},
  \bibinfo {author} {\bibfnamefont {J.~G.}\ \bibnamefont {Analytis}}, \bibinfo
  {author} {\bibfnamefont {Y.~L.}\ \bibnamefont {Chen}}, \bibinfo {author}
  {\bibfnamefont {I.~R.}\ \bibnamefont {Fisher}}, \bibinfo {author}
  {\bibfnamefont {P.~S.}\ \bibnamefont {Kirchmann}}, \ and\ \bibinfo {author}
  {\bibfnamefont {Z.-X.}\ \bibnamefont {Shen}},\ }\href {\doibase
  10.1103/PhysRevLett.108.117403} {\bibfield  {journal} {\bibinfo  {journal}
  {Phys. Rev. Lett.}\ }\textbf {\bibinfo {volume} {108}},\ \bibinfo {pages}
  {117403} (\bibinfo {year} {2012})}\BibitemShut {NoStop}%
\bibitem [{\citenamefont {Sobota}\ \emph {et~al.}(2013)\citenamefont {Sobota},
  \citenamefont {Yang}, \citenamefont {Kemper}, \citenamefont {Lee},
  \citenamefont {Schmitt}, \citenamefont {Li}, \citenamefont {Moore},
  \citenamefont {Analytis}, \citenamefont {Fisher}, \citenamefont {Kirchmann},
  \citenamefont {Devereaux},\ and\ \citenamefont {Shen}}]{Sobota2013}%
  \BibitemOpen
  \bibfield  {author} {\bibinfo {author} {\bibfnamefont {J.~A.}\ \bibnamefont
  {Sobota}}, \bibinfo {author} {\bibfnamefont {S.-L.}\ \bibnamefont {Yang}},
  \bibinfo {author} {\bibfnamefont {A.~F.}\ \bibnamefont {Kemper}}, \bibinfo
  {author} {\bibfnamefont {J.~J.}\ \bibnamefont {Lee}}, \bibinfo {author}
  {\bibfnamefont {F.~T.}\ \bibnamefont {Schmitt}}, \bibinfo {author}
  {\bibfnamefont {W.}~\bibnamefont {Li}}, \bibinfo {author} {\bibfnamefont
  {R.~G.}\ \bibnamefont {Moore}}, \bibinfo {author} {\bibfnamefont {J.~G.}\
  \bibnamefont {Analytis}}, \bibinfo {author} {\bibfnamefont {I.~R.}\
  \bibnamefont {Fisher}}, \bibinfo {author} {\bibfnamefont {P.~S.}\
  \bibnamefont {Kirchmann}}, \bibinfo {author} {\bibfnamefont {T.~P.}\
  \bibnamefont {Devereaux}}, \ and\ \bibinfo {author} {\bibfnamefont {Z.-X.}\
  \bibnamefont {Shen}},\ }\href {\doibase 10.1103/PhysRevLett.111.136802}
  {\bibfield  {journal} {\bibinfo  {journal} {Phys. Rev. Lett.}\ }\textbf
  {\bibinfo {volume} {111}},\ \bibinfo {pages} {136802} (\bibinfo {year}
  {2013})}\BibitemShut {NoStop}%
\bibitem [{\citenamefont {Niesner}\ \emph {et~al.}(2012)\citenamefont
  {Niesner}, \citenamefont {Fauster}, \citenamefont {Eremeev}, \citenamefont
  {Menshchikova}, \citenamefont {Koroteev}, \citenamefont {Protogenov},
  \citenamefont {Chulkov}, \citenamefont {Tereshchenko}, \citenamefont {Kokh},
  \citenamefont {Alekperov}, \citenamefont {Nadjafov},\ and\ \citenamefont
  {Mamedov}}]{Niesner2012}%
  \BibitemOpen
  \bibfield  {author} {\bibinfo {author} {\bibfnamefont {D.}~\bibnamefont
  {Niesner}}, \bibinfo {author} {\bibfnamefont {T.}~\bibnamefont {Fauster}},
  \bibinfo {author} {\bibfnamefont {S.~V.}\ \bibnamefont {Eremeev}}, \bibinfo
  {author} {\bibfnamefont {T.~V.}\ \bibnamefont {Menshchikova}}, \bibinfo
  {author} {\bibfnamefont {Y.~M.}\ \bibnamefont {Koroteev}}, \bibinfo {author}
  {\bibfnamefont {A.~P.}\ \bibnamefont {Protogenov}}, \bibinfo {author}
  {\bibfnamefont {E.~V.}\ \bibnamefont {Chulkov}}, \bibinfo {author}
  {\bibfnamefont {O.~E.}\ \bibnamefont {Tereshchenko}}, \bibinfo {author}
  {\bibfnamefont {K.~A.}\ \bibnamefont {Kokh}}, \bibinfo {author}
  {\bibfnamefont {O.}~\bibnamefont {Alekperov}}, \bibinfo {author}
  {\bibfnamefont {A.}~\bibnamefont {Nadjafov}}, \ and\ \bibinfo {author}
  {\bibfnamefont {N.}~\bibnamefont {Mamedov}},\ }\href {\doibase
  10.1103/PhysRevB.86.205403} {\bibfield  {journal} {\bibinfo  {journal} {Phys.
  Rev. B}\ }\textbf {\bibinfo {volume} {86}},\ \bibinfo {pages} {205403}
  (\bibinfo {year} {2012})}\BibitemShut {NoStop}%
\bibitem [{\citenamefont {Niesner}\ \emph
  {et~al.}(2014{\natexlab{a}})\citenamefont {Niesner}, \citenamefont {Otto},
  \citenamefont {Hermann}, \citenamefont {Fauster}, \citenamefont
  {Menshchikova}, \citenamefont {Eremeev}, \citenamefont {Aliev}, \citenamefont
  {Amiraslanov}, \citenamefont {Babanly}, \citenamefont {Echenique},\ and\
  \citenamefont {Chulkov}}]{Niesner2014}%
  \BibitemOpen
  \bibfield  {author} {\bibinfo {author} {\bibfnamefont {D.}~\bibnamefont
  {Niesner}}, \bibinfo {author} {\bibfnamefont {S.}~\bibnamefont {Otto}},
  \bibinfo {author} {\bibfnamefont {V.}~\bibnamefont {Hermann}}, \bibinfo
  {author} {\bibfnamefont {T.}~\bibnamefont {Fauster}}, \bibinfo {author}
  {\bibfnamefont {T.~V.}\ \bibnamefont {Menshchikova}}, \bibinfo {author}
  {\bibfnamefont {S.~V.}\ \bibnamefont {Eremeev}}, \bibinfo {author}
  {\bibfnamefont {Z.~S.}\ \bibnamefont {Aliev}}, \bibinfo {author}
  {\bibfnamefont {I.~R.}\ \bibnamefont {Amiraslanov}}, \bibinfo {author}
  {\bibfnamefont {M.~B.}\ \bibnamefont {Babanly}}, \bibinfo {author}
  {\bibfnamefont {P.~M.}\ \bibnamefont {Echenique}}, \ and\ \bibinfo {author}
  {\bibfnamefont {E.~V.}\ \bibnamefont {Chulkov}},\ }\href {\doibase
  10.1103/PhysRevB.89.081404} {\bibfield  {journal} {\bibinfo  {journal} {Phys.
  Rev. B}\ }\textbf {\bibinfo {volume} {89}},\ \bibinfo {pages} {081404}
  (\bibinfo {year} {2014}{\natexlab{a}})}\BibitemShut {NoStop}%
\bibitem [{\citenamefont {Niesner}\ \emph
  {et~al.}(2014{\natexlab{b}})\citenamefont {Niesner}, \citenamefont {Otto},
  \citenamefont {Fauster}, \citenamefont {Chulkov}, \citenamefont {Eremeev},
  \citenamefont {Tereshchenko},\ and\ \citenamefont {Kokh}}]{Niesner2014B}%
  \BibitemOpen
  \bibfield  {author} {\bibinfo {author} {\bibfnamefont {D.}~\bibnamefont
  {Niesner}}, \bibinfo {author} {\bibfnamefont {S.}~\bibnamefont {Otto}},
  \bibinfo {author} {\bibfnamefont {T.}~\bibnamefont {Fauster}}, \bibinfo
  {author} {\bibfnamefont {E.}~\bibnamefont {Chulkov}}, \bibinfo {author}
  {\bibfnamefont {S.}~\bibnamefont {Eremeev}}, \bibinfo {author} {\bibfnamefont
  {O.}~\bibnamefont {Tereshchenko}}, \ and\ \bibinfo {author} {\bibfnamefont
  {K.}~\bibnamefont {Kokh}},\ }\href {\doibase
  https://doi.org/10.1016/j.elspec.2014.03.013} {\bibfield  {journal} {\bibinfo
   {journal} {Journal of Electron Spectroscopy and Related Phenomena}\ }\textbf
  {\bibinfo {volume} {195}},\ \bibinfo {pages} {258 } (\bibinfo {year}
  {2014}{\natexlab{b}})}\BibitemShut {NoStop}%
\bibitem [{\citenamefont {Gundlach}(1966)}]{Gundlach1966}%
  \BibitemOpen
  \bibfield  {author} {\bibinfo {author} {\bibfnamefont {K.~H.}\ \bibnamefont
  {Gundlach}},\ }\href {\doibase 10.1016/0038-1101(66)90071-2} {\bibfield
  {journal} {\bibinfo  {journal} {Solid-State Electronics}\ }\textbf {\bibinfo
  {volume} {9}},\ \bibinfo {pages} {949} (\bibinfo {year} {1966})}\BibitemShut
  {NoStop}%
\bibitem [{\citenamefont {H{\"{o}}fer}\ \emph {et~al.}(1997)\citenamefont
  {H{\"{o}}fer}, \citenamefont {Shumay}, \citenamefont {Reu{\ss}},
  \citenamefont {Thomann}, \citenamefont {Wallauer},\ and\ \citenamefont
  {Fauster}}]{Hofer1997}%
  \BibitemOpen
  \bibfield  {author} {\bibinfo {author} {\bibfnamefont {U.}~\bibnamefont
  {H{\"{o}}fer}}, \bibinfo {author} {\bibfnamefont {I.~L.}\ \bibnamefont
  {Shumay}}, \bibinfo {author} {\bibfnamefont {C.}~\bibnamefont {Reu{\ss}}},
  \bibinfo {author} {\bibfnamefont {U.}~\bibnamefont {Thomann}}, \bibinfo
  {author} {\bibfnamefont {W.}~\bibnamefont {Wallauer}}, \ and\ \bibinfo
  {author} {\bibfnamefont {T.}~\bibnamefont {Fauster}},\ }\href {\doibase
  10.1126/science.277.5331.1480} {\bibfield  {journal} {\bibinfo  {journal}
  {Science}\ }\textbf {\bibinfo {volume} {277}},\ \bibinfo {pages} {1480}
  (\bibinfo {year} {1997})}\BibitemShut {NoStop}%
\bibitem [{\citenamefont {Kisiel}\ \emph {et~al.}(2011)\citenamefont {Kisiel},
  \citenamefont {Gnecco}, \citenamefont {Gysin}, \citenamefont {Marot},
  \citenamefont {Rast},\ and\ \citenamefont {Meyer}}]{Kisiel2011}%
  \BibitemOpen
  \bibfield  {author} {\bibinfo {author} {\bibfnamefont {M.}~\bibnamefont
  {Kisiel}}, \bibinfo {author} {\bibfnamefont {E.}~\bibnamefont {Gnecco}},
  \bibinfo {author} {\bibfnamefont {U.}~\bibnamefont {Gysin}}, \bibinfo
  {author} {\bibfnamefont {L.}~\bibnamefont {Marot}}, \bibinfo {author}
  {\bibfnamefont {S.}~\bibnamefont {Rast}}, \ and\ \bibinfo {author}
  {\bibfnamefont {E.}~\bibnamefont {Meyer}},\ }\href
  {http://dx.doi.org/10.1038/nmat2936} {\bibfield  {journal} {\bibinfo
  {journal} {Nature Materials}\ }\textbf {\bibinfo {volume} {10}},\ \bibinfo
  {pages} {119 EP } (\bibinfo {year} {2011})}\BibitemShut {NoStop}%
\bibitem [{\citenamefont {Langer}\ \emph {et~al.}(2013)\citenamefont {Langer},
  \citenamefont {Kisiel}, \citenamefont {Pawlak}, \citenamefont {Pellegrini},
  \citenamefont {Santoro}, \citenamefont {Buzio}, \citenamefont {Gerbi},
  \citenamefont {Balakrishnan}, \citenamefont {Baratoff}, \citenamefont
  {Tosatti},\ and\ \citenamefont {Meyer}}]{Langer2013}%
  \BibitemOpen
  \bibfield  {author} {\bibinfo {author} {\bibfnamefont {M.}~\bibnamefont
  {Langer}}, \bibinfo {author} {\bibfnamefont {M.}~\bibnamefont {Kisiel}},
  \bibinfo {author} {\bibfnamefont {R.}~\bibnamefont {Pawlak}}, \bibinfo
  {author} {\bibfnamefont {F.}~\bibnamefont {Pellegrini}}, \bibinfo {author}
  {\bibfnamefont {G.~E.}\ \bibnamefont {Santoro}}, \bibinfo {author}
  {\bibfnamefont {R.}~\bibnamefont {Buzio}}, \bibinfo {author} {\bibfnamefont
  {A.}~\bibnamefont {Gerbi}}, \bibinfo {author} {\bibfnamefont
  {G.}~\bibnamefont {Balakrishnan}}, \bibinfo {author} {\bibfnamefont
  {A.}~\bibnamefont {Baratoff}}, \bibinfo {author} {\bibfnamefont
  {E.}~\bibnamefont {Tosatti}}, \ and\ \bibinfo {author} {\bibfnamefont
  {E.}~\bibnamefont {Meyer}},\ }\href {http://dx.doi.org/10.1038/nmat3836}
  {\bibfield  {journal} {\bibinfo  {journal} {Nature Materials}\ }\textbf
  {\bibinfo {volume} {13}},\ \bibinfo {pages} {173 EP } (\bibinfo {year}
  {2013})}\BibitemShut {NoStop}%
\bibitem [{\citenamefont {Kisiel}\ \emph {et~al.}(2015)\citenamefont {Kisiel},
  \citenamefont {Pellegrini}, \citenamefont {Santoro}, \citenamefont
  {Samadashvili}, \citenamefont {Pawlak}, \citenamefont {Benassi},
  \citenamefont {Gysin}, \citenamefont {Buzio}, \citenamefont {Gerbi},
  \citenamefont {Meyer},\ and\ \citenamefont {Tosatti}}]{Kisiel2015}%
  \BibitemOpen
  \bibfield  {author} {\bibinfo {author} {\bibfnamefont {M.}~\bibnamefont
  {Kisiel}}, \bibinfo {author} {\bibfnamefont {F.}~\bibnamefont {Pellegrini}},
  \bibinfo {author} {\bibfnamefont {G.~E.}\ \bibnamefont {Santoro}}, \bibinfo
  {author} {\bibfnamefont {M.}~\bibnamefont {Samadashvili}}, \bibinfo {author}
  {\bibfnamefont {R.}~\bibnamefont {Pawlak}}, \bibinfo {author} {\bibfnamefont
  {A.}~\bibnamefont {Benassi}}, \bibinfo {author} {\bibfnamefont
  {U.}~\bibnamefont {Gysin}}, \bibinfo {author} {\bibfnamefont
  {R.}~\bibnamefont {Buzio}}, \bibinfo {author} {\bibfnamefont
  {A.}~\bibnamefont {Gerbi}}, \bibinfo {author} {\bibfnamefont
  {E.}~\bibnamefont {Meyer}}, \ and\ \bibinfo {author} {\bibfnamefont
  {E.}~\bibnamefont {Tosatti}},\ }\href {\doibase
  10.1103/PhysRevLett.115.046101} {\bibfield  {journal} {\bibinfo  {journal}
  {Phys. Rev. Lett.}\ }\textbf {\bibinfo {volume} {115}},\ \bibinfo {pages}
  {046101} (\bibinfo {year} {2015})}\BibitemShut {NoStop}%
\bibitem [{\citenamefont {Stipe}\ \emph {et~al.}(2001)\citenamefont {Stipe},
  \citenamefont {Mamin}, \citenamefont {Stowe}, \citenamefont {Kenny},\ and\
  \citenamefont {Rugar}}]{Stipe2001}%
  \BibitemOpen
  \bibfield  {author} {\bibinfo {author} {\bibfnamefont {B.~C.}\ \bibnamefont
  {Stipe}}, \bibinfo {author} {\bibfnamefont {H.~J.}\ \bibnamefont {Mamin}},
  \bibinfo {author} {\bibfnamefont {T.~D.}\ \bibnamefont {Stowe}}, \bibinfo
  {author} {\bibfnamefont {T.~W.}\ \bibnamefont {Kenny}}, \ and\ \bibinfo
  {author} {\bibfnamefont {D.}~\bibnamefont {Rugar}},\ }\href {\doibase
  10.1103/PhysRevLett.87.096801} {\bibfield  {journal} {\bibinfo  {journal}
  {Phys. Rev. Lett.}\ }\textbf {\bibinfo {volume} {87}},\ \bibinfo {pages}
  {096801} (\bibinfo {year} {2001})}\BibitemShut {NoStop}%
\bibitem [{\citenamefont {Volokitin}\ and\ \citenamefont
  {Persson}(2007)}]{volokitin2007}%
  \BibitemOpen
  \bibfield  {author} {\bibinfo {author} {\bibfnamefont {A.~I.}\ \bibnamefont
  {Volokitin}}\ and\ \bibinfo {author} {\bibfnamefont {B.~N.~J.}\ \bibnamefont
  {Persson}},\ }\href {\doibase 10.1103/RevModPhys.79.1291} {\bibfield
  {journal} {\bibinfo  {journal} {Rev. Mod. Phys.}\ }\textbf {\bibinfo {volume}
  {79}},\ \bibinfo {pages} {1291} (\bibinfo {year} {2007})}\BibitemShut
  {NoStop}%
\bibitem [{\citenamefont {Su}\ \emph {et~al.}(2016)\citenamefont {Su},
  \citenamefont {Lin}, \citenamefont {Chan}, \citenamefont {Lu},\ and\
  \citenamefont {Chang}}]{WeiBin2016}%
  \BibitemOpen
  \bibfield  {author} {\bibinfo {author} {\bibfnamefont {W.-B.}\ \bibnamefont
  {Su}}, \bibinfo {author} {\bibfnamefont {C.-L.}\ \bibnamefont {Lin}},
  \bibinfo {author} {\bibfnamefont {W.-Y.}\ \bibnamefont {Chan}}, \bibinfo
  {author} {\bibfnamefont {S.-M.}\ \bibnamefont {Lu}}, \ and\ \bibinfo {author}
  {\bibfnamefont {C.-S.}\ \bibnamefont {Chang}},\ }\href
  {http://stacks.iop.org/0957-4484/27/i=17/a=175705} {\bibfield  {journal}
  {\bibinfo  {journal} {Nanotechnology}\ }\textbf {\bibinfo {volume} {27}},\
  \bibinfo {pages} {175705} (\bibinfo {year} {2016})}\BibitemShut {NoStop}%
\bibitem [{\citenamefont {Neupane}\ \emph {et~al.}(2012)\citenamefont
  {Neupane}, \citenamefont {Xu}, \citenamefont {Wray}, \citenamefont
  {Petersen}, \citenamefont {Shankar}, \citenamefont {Alidoust}, \citenamefont
  {Liu}, \citenamefont {Fedorov}, \citenamefont {Ji}, \citenamefont {Allred},
  \citenamefont {Hor}, \citenamefont {Chang}, \citenamefont {Jeng},
  \citenamefont {Lin}, \citenamefont {Bansil}, \citenamefont {Cava},\ and\
  \citenamefont {Hasan}}]{Neupane2012}%
  \BibitemOpen
  \bibfield  {author} {\bibinfo {author} {\bibfnamefont {M.}~\bibnamefont
  {Neupane}}, \bibinfo {author} {\bibfnamefont {S.-Y.}\ \bibnamefont {Xu}},
  \bibinfo {author} {\bibfnamefont {L.~A.}\ \bibnamefont {Wray}}, \bibinfo
  {author} {\bibfnamefont {A.}~\bibnamefont {Petersen}}, \bibinfo {author}
  {\bibfnamefont {R.}~\bibnamefont {Shankar}}, \bibinfo {author} {\bibfnamefont
  {N.}~\bibnamefont {Alidoust}}, \bibinfo {author} {\bibfnamefont
  {C.}~\bibnamefont {Liu}}, \bibinfo {author} {\bibfnamefont {A.}~\bibnamefont
  {Fedorov}}, \bibinfo {author} {\bibfnamefont {H.}~\bibnamefont {Ji}},
  \bibinfo {author} {\bibfnamefont {J.~M.}\ \bibnamefont {Allred}}, \bibinfo
  {author} {\bibfnamefont {Y.~S.}\ \bibnamefont {Hor}}, \bibinfo {author}
  {\bibfnamefont {T.-R.}\ \bibnamefont {Chang}}, \bibinfo {author}
  {\bibfnamefont {H.-T.}\ \bibnamefont {Jeng}}, \bibinfo {author}
  {\bibfnamefont {H.}~\bibnamefont {Lin}}, \bibinfo {author} {\bibfnamefont
  {A.}~\bibnamefont {Bansil}}, \bibinfo {author} {\bibfnamefont {R.~J.}\
  \bibnamefont {Cava}}, \ and\ \bibinfo {author} {\bibfnamefont {M.~Z.}\
  \bibnamefont {Hasan}},\ }\href {\doibase 10.1103/PhysRevB.85.235406}
  {\bibfield  {journal} {\bibinfo  {journal} {Phys. Rev. B}\ }\textbf {\bibinfo
  {volume} {85}},\ \bibinfo {pages} {235406} (\bibinfo {year}
  {2012})}\BibitemShut {NoStop}%
\bibitem [{\citenamefont {Miyamoto}\ \emph {et~al.}(2012)\citenamefont
  {Miyamoto}, \citenamefont {Kimura}, \citenamefont {Okuda}, \citenamefont
  {Miyahara}, \citenamefont {Kuroda}, \citenamefont {Namatame}, \citenamefont
  {Taniguchi}, \citenamefont {Eremeev}, \citenamefont {Menshchikova},
  \citenamefont {Chulkov}, \citenamefont {Kokh},\ and\ \citenamefont
  {Tereshchenko}}]{Miyamoto2012}%
  \BibitemOpen
  \bibfield  {author} {\bibinfo {author} {\bibfnamefont {K.}~\bibnamefont
  {Miyamoto}}, \bibinfo {author} {\bibfnamefont {A.}~\bibnamefont {Kimura}},
  \bibinfo {author} {\bibfnamefont {T.}~\bibnamefont {Okuda}}, \bibinfo
  {author} {\bibfnamefont {H.}~\bibnamefont {Miyahara}}, \bibinfo {author}
  {\bibfnamefont {K.}~\bibnamefont {Kuroda}}, \bibinfo {author} {\bibfnamefont
  {H.}~\bibnamefont {Namatame}}, \bibinfo {author} {\bibfnamefont
  {M.}~\bibnamefont {Taniguchi}}, \bibinfo {author} {\bibfnamefont {S.~V.}\
  \bibnamefont {Eremeev}}, \bibinfo {author} {\bibfnamefont {T.~V.}\
  \bibnamefont {Menshchikova}}, \bibinfo {author} {\bibfnamefont {E.~V.}\
  \bibnamefont {Chulkov}}, \bibinfo {author} {\bibfnamefont {K.~A.}\
  \bibnamefont {Kokh}}, \ and\ \bibinfo {author} {\bibfnamefont {O.~E.}\
  \bibnamefont {Tereshchenko}},\ }\href {\doibase
  10.1103/PhysRevLett.109.166802} {\bibfield  {journal} {\bibinfo  {journal}
  {Phys. Rev. Lett.}\ }\textbf {\bibinfo {volume} {109}},\ \bibinfo {pages}
  {166802} (\bibinfo {year} {2012})}\BibitemShut {NoStop}%
\bibitem [{\citenamefont {Schouteden}\ \emph {et~al.}(2016)\citenamefont
  {Schouteden}, \citenamefont {Li}, \citenamefont {Chen}, \citenamefont {Song},
  \citenamefont {Partoens}, \citenamefont {Van~Haesendonck},\ and\
  \citenamefont {Park}}]{Schouteden2016}%
  \BibitemOpen
  \bibfield  {author} {\bibinfo {author} {\bibfnamefont {K.}~\bibnamefont
  {Schouteden}}, \bibinfo {author} {\bibfnamefont {Z.}~\bibnamefont {Li}},
  \bibinfo {author} {\bibfnamefont {T.}~\bibnamefont {Chen}}, \bibinfo {author}
  {\bibfnamefont {F.}~\bibnamefont {Song}}, \bibinfo {author} {\bibfnamefont
  {B.}~\bibnamefont {Partoens}}, \bibinfo {author} {\bibfnamefont
  {C.}~\bibnamefont {Van~Haesendonck}}, \ and\ \bibinfo {author} {\bibfnamefont
  {K.}~\bibnamefont {Park}},\ }\href {http://dx.doi.org/10.1038/srep20278}
  {\bibfield  {journal} {\bibinfo  {journal} {Scientific Reports}\ }\textbf
  {\bibinfo {volume} {6}},\ \bibinfo {pages} {20278 EP } (\bibinfo {year}
  {2016})}\BibitemShut {NoStop}%
\bibitem [{\citenamefont {Pivetta}\ \emph {et~al.}(2005)\citenamefont
  {Pivetta}, \citenamefont {Patthey}, \citenamefont {Stengel}, \citenamefont
  {Baldereschi},\ and\ \citenamefont {Schneider}}]{Pivetta2005}%
  \BibitemOpen
  \bibfield  {author} {\bibinfo {author} {\bibfnamefont {M.}~\bibnamefont
  {Pivetta}}, \bibinfo {author} {\bibfnamefont {F.}~\bibnamefont {Patthey}},
  \bibinfo {author} {\bibfnamefont {M.}~\bibnamefont {Stengel}}, \bibinfo
  {author} {\bibfnamefont {A.}~\bibnamefont {Baldereschi}}, \ and\ \bibinfo
  {author} {\bibfnamefont {W.~D.}\ \bibnamefont {Schneider}},\ }\href {\doibase
  10.1103/PhysRevB.72.115404} {\bibfield  {journal} {\bibinfo  {journal}
  {Physical Review B - Condensed Matter and Materials Physics}\ }\textbf
  {\bibinfo {volume} {72}},\ \bibinfo {pages} {1} (\bibinfo {year}
  {2005})}\BibitemShut {NoStop}%
\bibitem [{\citenamefont {Bose}\ \emph {et~al.}(2010)\citenamefont {Bose},
  \citenamefont {Silkin}, \citenamefont {Ohmann}, \citenamefont {Brihuega},
  \citenamefont {Vitali}, \citenamefont {Michaelis}, \citenamefont {Mallet},
  \citenamefont {Veuillen}, \citenamefont {{Alexander Schneider}},
  \citenamefont {Chulkov}, \citenamefont {Echenique},\ and\ \citenamefont
  {Kern}}]{Bose2010}%
  \BibitemOpen
  \bibfield  {author} {\bibinfo {author} {\bibfnamefont {S.}~\bibnamefont
  {Bose}}, \bibinfo {author} {\bibfnamefont {V.~M.}\ \bibnamefont {Silkin}},
  \bibinfo {author} {\bibfnamefont {R.}~\bibnamefont {Ohmann}}, \bibinfo
  {author} {\bibfnamefont {I.}~\bibnamefont {Brihuega}}, \bibinfo {author}
  {\bibfnamefont {L.}~\bibnamefont {Vitali}}, \bibinfo {author} {\bibfnamefont
  {C.~H.}\ \bibnamefont {Michaelis}}, \bibinfo {author} {\bibfnamefont
  {P.}~\bibnamefont {Mallet}}, \bibinfo {author} {\bibfnamefont {J.~Y.}\
  \bibnamefont {Veuillen}}, \bibinfo {author} {\bibfnamefont {M.}~\bibnamefont
  {{Alexander Schneider}}}, \bibinfo {author} {\bibfnamefont {E.~V.}\
  \bibnamefont {Chulkov}}, \bibinfo {author} {\bibfnamefont {P.~M.}\
  \bibnamefont {Echenique}}, \ and\ \bibinfo {author} {\bibfnamefont
  {K.}~\bibnamefont {Kern}},\ }\href {\doibase 10.1088/1367-2630/12/2/023028}
  {\bibfield  {journal} {\bibinfo  {journal} {New Journal of Physics}\ }\textbf
  {\bibinfo {volume} {12}} (\bibinfo {year} {2010})}\  \BibitemShut {NoStop}%
\bibitem [{\citenamefont {Volokitin}\ \emph {et~al.}(2006)\citenamefont
  {Volokitin}, \citenamefont {Persson},\ and\ \citenamefont
  {Ueba}}]{Volokitin2006}%
  \BibitemOpen
  \bibfield  {author} {\bibinfo {author} {\bibfnamefont {A.~I.}\ \bibnamefont
  {Volokitin}}, \bibinfo {author} {\bibfnamefont {B.~N.~J.}\ \bibnamefont
  {Persson}}, \ and\ \bibinfo {author} {\bibfnamefont {H.}~\bibnamefont
  {Ueba}},\ }\href {\doibase 10.1103/PhysRevB.73.165423} {\bibfield  {journal}
  {\bibinfo  {journal} {Physical Review B}\ }\textbf {\bibinfo {volume} {73}},\
  \bibinfo {pages} {165423} (\bibinfo {year} {2006})}\BibitemShut {NoStop}%
\bibitem [{\citenamefont {Cockins}\ \emph {et~al.}(2010)\citenamefont
  {Cockins}, \citenamefont {Miyahara}, \citenamefont {Bennett}, \citenamefont
  {Clerk}, \citenamefont {Studenikin}, \citenamefont {Poole}, \citenamefont
  {Sachrajda},\ and\ \citenamefont {Grutter}}]{Cockins2010}%
  \BibitemOpen
  \bibfield  {author} {\bibinfo {author} {\bibfnamefont {L.}~\bibnamefont
  {Cockins}}, \bibinfo {author} {\bibfnamefont {Y.}~\bibnamefont {Miyahara}},
  \bibinfo {author} {\bibfnamefont {S.~D.}\ \bibnamefont {Bennett}}, \bibinfo
  {author} {\bibfnamefont {A.~A.}\ \bibnamefont {Clerk}}, \bibinfo {author}
  {\bibfnamefont {S.}~\bibnamefont {Studenikin}}, \bibinfo {author}
  {\bibfnamefont {P.}~\bibnamefont {Poole}}, \bibinfo {author} {\bibfnamefont
  {A.}~\bibnamefont {Sachrajda}}, \ and\ \bibinfo {author} {\bibfnamefont
  {P.}~\bibnamefont {Grutter}},\ }\href {\doibase 10.1073/pnas.0912716107}
  {\bibfield  {journal} {\bibinfo  {journal} {Proceedings of the National
  Academy of Sciences}\ }\textbf {\bibinfo {volume} {107}},\ \bibinfo {pages}
  {9496} (\bibinfo {year} {2010})}\BibitemShut {NoStop}%
\bibitem [{\citenamefont {Kisiel}\ \emph {et~al.}(2018)\citenamefont {Kisiel},
  \citenamefont {Brovko}, \citenamefont {Yildiz}, \citenamefont {Pawlak},
  \citenamefont {Gysin}, \citenamefont {Tosatti},\ and\ \citenamefont
  {Meyer}}]{Kisiel2018}%
  \BibitemOpen
  \bibfield  {author} {\bibinfo {author} {\bibfnamefont {M.}~\bibnamefont
  {Kisiel}}, \bibinfo {author} {\bibfnamefont {O.~O.}\ \bibnamefont {Brovko}},
  \bibinfo {author} {\bibfnamefont {D.}~\bibnamefont {Yildiz}}, \bibinfo
  {author} {\bibfnamefont {R.}~\bibnamefont {Pawlak}}, \bibinfo {author}
  {\bibfnamefont {U.}~\bibnamefont {Gysin}}, \bibinfo {author} {\bibfnamefont
  {E.}~\bibnamefont {Tosatti}}, \ and\ \bibinfo {author} {\bibfnamefont
  {E.}~\bibnamefont {Meyer}},\ }\href {\doibase 10.1038/s41467-018-05392-1}
  {\bibfield  {journal} {\bibinfo  {journal} {Nature Communications}\ }\textbf
  {\bibinfo {volume} {9}},\ \bibinfo {pages} {2946} (\bibinfo {year}
  {2018})}\BibitemShut {NoStop}%
\bibitem [{\citenamefont {Gnecco}\ and\ \citenamefont
  {Meyer}(2007)}]{gnecco_meyer_2007}%
  \BibitemOpen
  \bibfield  {author} {\bibinfo {author} {\bibfnamefont {E.}~\bibnamefont
  {Gnecco}}\ and\ \bibinfo {author} {\bibfnamefont {E.}~\bibnamefont {Meyer}},\
  }\href@noop {} {\emph {\bibinfo {title} {Fundamentals of friction and wear on
  the nanoscale}}}\ (\bibinfo  {publisher} {Springer-Verlag},\ \bibinfo {year}
  {2007})\ p.\ \bibinfo {pages} {426}\BibitemShut {NoStop}%
\bibitem [{\citenamefont {Sessi}\ \emph {et~al.}(2014)\citenamefont {Sessi},
  \citenamefont {Reis}, \citenamefont {Bathon}, \citenamefont {Kokh},
  \citenamefont {Tereshchenko},\ and\ \citenamefont {Bode}}]{Sessi2014}%
  \BibitemOpen
  \bibfield  {author} {\bibinfo {author} {\bibfnamefont {P.}~\bibnamefont
  {Sessi}}, \bibinfo {author} {\bibfnamefont {F.}~\bibnamefont {Reis}},
  \bibinfo {author} {\bibfnamefont {T.}~\bibnamefont {Bathon}}, \bibinfo
  {author} {\bibfnamefont {K.~A.}\ \bibnamefont {Kokh}}, \bibinfo {author}
  {\bibfnamefont {O.~E.}\ \bibnamefont {Tereshchenko}}, \ and\ \bibinfo
  {author} {\bibfnamefont {M.}~\bibnamefont {Bode}},\ }\href
  {http://dx.doi.org/10.1038/ncomms6349} {\bibfield  {journal} {\bibinfo
  {journal} {Nature Communications}\ }\textbf {\bibinfo {volume} {5}},\
  \bibinfo {pages} {5349 EP } (\bibinfo {year} {2014})}\BibitemShut {NoStop}%
\bibitem [{\citenamefont {Olsen}(1962)}]{olsen_1962}%
  \BibitemOpen
  \bibfield  {author} {\bibinfo {author} {\bibfnamefont {J.~L.}\ \bibnamefont
  {Olsen}},\ }\href@noop {} {\emph {\bibinfo {title} {Electron transport in
  metals}}}\ (\bibinfo  {publisher} {Interscience},\ \bibinfo {year}
  {1962})\BibitemShut {NoStop}%
\bibitem [{\citenamefont {Park}\ \emph {et~al.}(2016)\citenamefont {Park},
  \citenamefont {Park}, \citenamefont {Jeong}, \citenamefont {Jeong},
  \citenamefont {Song},\ and\ \citenamefont {Cho}}]{Park2016}%
  \BibitemOpen
  \bibfield  {author} {\bibinfo {author} {\bibfnamefont {D.}~\bibnamefont
  {Park}}, \bibinfo {author} {\bibfnamefont {S.}~\bibnamefont {Park}}, \bibinfo
  {author} {\bibfnamefont {K.}~\bibnamefont {Jeong}}, \bibinfo {author}
  {\bibfnamefont {H.-S.}\ \bibnamefont {Jeong}}, \bibinfo {author}
  {\bibfnamefont {J.~Y.}\ \bibnamefont {Song}}, \ and\ \bibinfo {author}
  {\bibfnamefont {M.-H.}\ \bibnamefont {Cho}},\ }\href
  {http://dx.doi.org/10.1038/srep19132} {\bibfield  {journal} {\bibinfo
  {journal} {Scientific Reports}\ }\textbf {\bibinfo {volume} {6}},\ \bibinfo
  {pages} {19132 EP } (\bibinfo {year} {2016})}\BibitemShut {NoStop}%
\bibitem [{\citenamefont {Stomp}\ \emph {et~al.}(2005)\citenamefont {Stomp},
  \citenamefont {Miyahara}, \citenamefont {Schaer}, \citenamefont {Sun},
  \citenamefont {Guo}, \citenamefont {Grutter}, \citenamefont {Studenikin},
  \citenamefont {Poole},\ and\ \citenamefont {Sachrajda}}]{Stomp2005}%
  \BibitemOpen
  \bibfield  {author} {\bibinfo {author} {\bibfnamefont {R.}~\bibnamefont
  {Stomp}}, \bibinfo {author} {\bibfnamefont {Y.}~\bibnamefont {Miyahara}},
  \bibinfo {author} {\bibfnamefont {S.}~\bibnamefont {Schaer}}, \bibinfo
  {author} {\bibfnamefont {Q.}~\bibnamefont {Sun}}, \bibinfo {author}
  {\bibfnamefont {H.}~\bibnamefont {Guo}}, \bibinfo {author} {\bibfnamefont
  {P.}~\bibnamefont {Grutter}}, \bibinfo {author} {\bibfnamefont
  {S.}~\bibnamefont {Studenikin}}, \bibinfo {author} {\bibfnamefont
  {P.}~\bibnamefont {Poole}}, \ and\ \bibinfo {author} {\bibfnamefont
  {A.}~\bibnamefont {Sachrajda}},\ }\href {\doibase
  10.1103/PhysRevLett.94.056802} {\bibfield  {journal} {\bibinfo  {journal}
  {Phys. Rev. Lett.}\ }\textbf {\bibinfo {volume} {94}},\ \bibinfo {pages}
  {056802} (\bibinfo {year} {2005})}\BibitemShut {NoStop}%
\bibitem [{\citenamefont {Cleveland}\ \emph {et~al.}(1998)\citenamefont
  {Cleveland}, \citenamefont {Anczykowski}, \citenamefont {Schmid},\ and\
  \citenamefont {Elings}}]{Cleveland1998}%
  \BibitemOpen
  \bibfield  {author} {\bibinfo {author} {\bibfnamefont {J.~P.}\ \bibnamefont
  {Cleveland}}, \bibinfo {author} {\bibfnamefont {B.}~\bibnamefont
  {Anczykowski}}, \bibinfo {author} {\bibfnamefont {a.~E.}\ \bibnamefont
  {Schmid}}, \ and\ \bibinfo {author} {\bibfnamefont {V.~B.}\ \bibnamefont
  {Elings}},\ }\href {\doibase 10.1063/1.121434} {\bibfield  {journal}
  {\bibinfo  {journal} {Applied Physics Letters}\ }\textbf {\bibinfo {volume}
  {72}},\ \bibinfo {pages} {2613} (\bibinfo {year} {1998}).}\BibitemShut
  {NoStop}%
  \bibitem [{\citenamefont {Fatayer}\ \emph {et~al.}(2018)\citenamefont {Fatayer},
  	\citenamefont {Moll}, \citenamefont {Collazos}, \citenamefont {P{\'{e}}rez},
  	\citenamefont {Guiti{\'{a}}n}, \citenamefont {Pe{\~{n}}a},
  	\citenamefont {Gross},\ and\ \citenamefont {Meyer}}]{Fatayer2018}%
  \BibitemOpen
  \bibfield  {author} {\bibinfo {author} {\bibfnamefont {S.}~\bibnamefont
  		{Fatayer}}, \bibinfo {author} {\bibfnamefont {N.}~\bibnamefont {Moll}},
  	\bibinfo {author} {\bibfnamefont {S.}~\bibnamefont {Collazos}}, \bibinfo
  	{author} {\bibfnamefont {D.}~\bibnamefont {P{\'{e}}rez}}, \bibinfo {author}
  	{\bibfnamefont {E.}~\bibnamefont {Guiti{\'{a}}n}}, \bibinfo {author} {\bibfnamefont
  		{D.}~\bibnamefont {Pe{\~{n}}a}}, \bibinfo {author} {\bibfnamefont
  		{L.}~\bibnamefont {Gross}}, \ and\ \bibinfo {author} {\bibfnamefont
  		{G.}~\bibnamefont {Meyer}},\ }\href {\doibase
  	10.1103/PhysRevLett.121.226101} {\bibfield  {journal} {\bibinfo  {journal}
  		{Phys. Rev. Lett.}\ }\textbf {\bibinfo {volume} {121}},\ \bibinfo {pages}
  	{226101} (\bibinfo {year} {2018}).}\BibitemShut {NoStop}%
  \bibitem [{\citenamefont {Paulsson}(2002)}]{Paulsson2002}%
  \BibitemOpen
  \bibfield  {author} {\bibinfo {author} {\bibfnamefont {M.}~\bibnamefont
  		{Paulsson}},\ \bibinfo {author} {\bibfnamefont {F.}~\bibnamefont {Zahid}},\ and\ \bibinfo {author} {\bibfnamefont {S.}~\bibnamefont {Datta},}\
  }\href@noop {} {\emph {\bibinfo {title} {Resistance of a Molecule}}}\ (\bibinfo  {publisher} {CRC Press},\ \bibinfo {year}
{2002})\ p.\ \bibinfo {pages} {426}.\BibitemShut {NoStop}%
  \bibitem [{\citenamefont {Sarid}(1991)}]{Sarid1991}%
  \BibitemOpen
  \bibfield  {author} {\bibinfo {author} {\bibfnamefont {D.}~\bibnamefont
  		{Sarid},}\
  }\href@noop {} {\emph {\bibinfo {title} {Scanning force microscopy with applications to electric, magnetic and atomic forces}}}\ (\bibinfo  {publisher} {Oxford University Press},\ \bibinfo {year}
  {1991}). \BibitemShut {NoStop}%
  




\end{thebibliography}
\end{document}